\documentclass[review,3p,times,12pt]{elsarticle}

\usepackage{hyperref}
\usepackage{graphicx}
\usepackage{amsmath}
\usepackage{xfrac}
\usepackage{color}

\DeclareMathOperator{\sech}{sech}

\journal{Journal of \LaTeX\ Templates}

\begin{document}

\begin{frontmatter}

\title{Two-way coupling of magnetohydrodynamic simulations with embedded particle-in-cell simulations}
\tnotetext[mytitlenote]{Fully documented templates are available in the elsarticle package on \href{http://www.ctan.org/tex-archive/macros/latex/contrib/elsarticle}{CTAN}.}

\author{K. D. Makwana, R. Keppens, G. Lapenta}
\address{Center for mathematical Plasma Astrophysics, Department of Mathematics, KU Leuven, Belgium}


%

\begin{abstract}
We describe a method for coupling an embedded domain in a magnetohydrodynamic (MHD) simulation with a particle-in-cell (PIC) method. In this two-way coupling {we follow the work of Daldorff et al.,~\cite{DaldorffToth2014}} in which the PIC domain receives its initial and boundary conditions from MHD variables (MHD to PIC coupling) while the MHD simulation is updated based on the PIC variables (PIC to MHD coupling). This method can be useful for simulating large plasma systems, where kinetic effects captured by particle-in-cell simulations are localized but affect global dynamics.  We describe the numerical implementation of this coupling, its time-stepping algorithm, and its parallelization strategy{, emphasizing the novel aspects of it}. We test the stability and energy/momentum conservation of this method by simulating a steady-state plasma. We test the dynamics of this coupling by propagating plasma waves through the embedded PIC domain. Coupling with MHD shows satisfactory results for the fast magnetosonic wave, but significant distortion for the circularly polarized Alfv{\'e}n wave. Coupling with Hall-MHD shows excellent coupling for the whistler wave. We also apply this methodology to simulate a Geospace Environmental Modeling (GEM) challenge type of reconnection with the diffusion region simulated by PIC coupled to larger scales with MHD and Hall-MHD. In both these cases we see the expected signatures of kinetic reconnection in the PIC domain, implying that this method can be used for reconnection studies.   
\end{abstract}

\begin{keyword}
MHD, PIC, space plasmas, multi-scale
\end{keyword}

\end{frontmatter}


\section{Introduction}


Plasmas exhibit a vast range of scales and a rich variety of phenomena. Plasma systems range in scales from table-top laser-plasma experiments~\cite{MalkaFaure2008} to intra-{galaxy} cluster medium~\cite{ZhuravlevaChurazov2014}. At the same time, different plasma phenomena occur at different length and time scales. At large scales, plasma acts like a perfectly conducting single fluid, which is suitably described by the ideal MHD model~\cite{Davidson}. At smaller scales we see differences between the electrons and ions, and this can be treated with more detailed models like the Hall-MHD or a two-fluid model. Magnetic reconnection is a ubiquitous phenomenon occurring in almost every kind of plasma system. In magnetic reconnection, the electron and ion motion separates below the ion skin-depth scale, giving rise to fast reconnection~\cite{GEM,ZweibelYamada2009}. However, the coupling of ion skin depth scale current sheets with large scale macroscopic current sheets is still not understood~\cite{TreumannBaumjohann2013}. Magnetic reconnection plays an important role in solar coronal flares. The coronal loop lengths are of the order of $10^8-10^9m$~\cite{Aschwanden2004}, the current sheet thickness in corona is of $\sim10^6-10^7m$~\cite{ZhuLiu2016}, whereas the ion skin-depth in corona is $\sim 10m$ and the electron skin depth is $\sim 10^{-1}m$~\cite{Tsiklauri2009}. Plasma turbulence is a similar multi-scale phenomena which can span over more than 10 orders of magnitude~\cite{AlexandrovaLacombe2012, ArmstrongRickett1995}. It is also a multi-physics phenomena as the nature of turbulent cascade changes at the ion and electron length scales~\cite{KiyaniOsman2015, MakwanaZhdankin2015}. 


Numerical simulation of this entire range of scales is not possible with present computational resources, and also does not seem possible anytime in the foreseeable future. Therefore, suitable numerical schemes have to be developed to accurately simulate such plasmas. Adaptive-mesh-refinement (AMR)~\cite{BergerCollela1989} can overcome some of this disparity in scales and several MHD codes include this capability. However, AMR can neither cover the entire range of scales nor the different physics at different scales. Even with AMR, MHD codes cannot capture the real physics at work in small scales. For this, a kinetic simulation, like particle-in-cell (PIC), is required. However, since kinetic simulations track a more complex phase space compared to fluid simulations, they are more computationally intensive. Therefore, it will not be possible to simulate such systems kinetically in their entirety in the foreseeable future.


In several physical systems kinetic effects can be localized. For example, collisionless shocks are mediated by electric and magnetic fields which require important kinetic effects to be present in and around the shock front~\cite{StockemNovoBret2016}. In electrostatic shocks, electric fields trap particles leading to kinetic effects. In magnetized shocks, the shock thickness is set by the ion gyro-radius and by the kinetic scale instabilities that occur in the shock interface region, whereas outside the plasma behaves like a fluid. In magnetic reconnection, the kinetic effects are very important in the narrow diffusion region where magnetic field lines break and reconnect. However, this change of topology can affect the dynamics of the entire system, like localized reconnection in the Earth's magnetosphere drives the entire space-weather system, or localized reconnection in solar corona can drive coronal mass ejections (CMEs) which affect the entire heliosphere. 



To attack such multi-scale, multi-physics problems, often times multiple codes are used in a coupled manner~\cite{TothHolst2012, RiekeTrost2015, SugiyamaKusano2007}. If we kinetically simulate only a localized region where kinetic physics is important, and simulate the rest of the global plasma with an MHD simulation, we can hope to achieve global simulations of such large scale systems with the required kinetic physics. This is a spatial coupling of MHD and PIC simulations. Such a coupling has been implemented and utilized for magnetosphere studies till now~\cite{DaldorffToth2014,TothJia2016}. That work coupled the BATS-R-US MHD code and the iPIC3D kinetic code~\cite{MarkidisLapenta2010}. In this work we couple iPIC3D with the MPI-AMRVAC code~\cite{PorthXia2014}, { using a simpler isotropic pressure tensor in MHD}. There are also important differences in the coupling methodology which we will outline below. In Sec. 2 we describe the methodology of this coupling involving the implementation of initial conditions, boundary conditions, time-stepping method and parallelization. In Sec. 3 we show tests of this coupling with the simulation of plasma waves, namely Alfv{\'e}n waves, whistler waves, and magnetosonic waves. In Sec. 4 we use this for simulating the Geospace Environmental Modeling (GEM) challenge ~\cite{GEM} problem which tests the applicability of this method to reconnection problems. We conclude with summary and discussion in Sec. 5.

\section{Coupling methodology}

The coupling is implemented between the codes MPI-AMRVAC~\cite{PorthXia2014} and iPIC3D~\cite{MarkidisLapenta2010, Lapenta2012}. MPI-AMRVAC solves a variety of hyperbolic advection problems, including MHD, with a variety of finite volume and finite difference schemes. It offers versatility in terms of geometry of the grid, dimensionality, the spatial and temporal discretization and computer platform~\cite{KeppensNool2003,Toth1997}. It also has resistive MHD and Hall-MHD modules which are critical in coupling with a kinetic code. It has a grid-block based adaptive mesh refinement scheme, where user can select the number of levels of refinement, with successive levels refined by a factor of 2. In this work, we are using MPI-AMRVAC with a fixed, uniform, domain-decomposed grid. However the AMR will be very useful for future applications. iPIC3D is a semi-implicit moment method for PIC simulations of plasmas. It { uses a semi-implicit scheme to solve for the electric field} by using a linearized version of the Amp{\`e}re's law. Then the magnetic field is obtained from the induction equation and the particles are moved implicitly. As a result of this semi-implicit scheme, iPIC3D can take larger grid spacing ($10\times-50\times$) and larger time steps ($5\times-10\times$) compared to explicit PIC codes. This makes it suitable for coupling with an MHD code like MPI-AMRVAC. Below we describe the method for a two-way coupling of these codes. In MHD to PIC coupling, information from MPI-AMRVAC is used to provide initial conditions and boundary conditions for iPIC3D. In PIC to MHD coupling information from iPIC3D is used to update the MPI-AMRVAC solution. This forms the two-way coupling method.

\subsection{MHD to PIC coupling}
\label{MHDtoPIC}
In this work, MPI-AMRVAC{~\cite{PorthXia2014,KeppensNool2003}} advances the conservative variables of mass density ($\rho$), momentum density ($\rho\mathbf{v}$), energy density ($e$), and magnetic field intensity ($\mathbf{B}$) according to the following MHD equations:
\begin{equation}
\frac{\partial \rho}{\partial t}+\nabla\cdot(\rho\mathbf{v})=S_{\rho},
\label{densityequation}
\end{equation}
\begin{equation}
\frac{\partial \rho\mathbf{v}}{\partial t}+\nabla\cdot(\mathbf{v}\rho\mathbf{v}-\mathbf{B}\mathbf{B})+\nabla p_{\mathrm{total}}=\mathbf{S}_{\rho v},
\label{momentumequation}
\end{equation}
\begin{equation}
\frac{\partial e}{\partial t}+\nabla\cdot(\mathbf{v}e+\mathbf{v}p_{\mathrm{total}}-\mathbf{B}\mathbf{B}\cdot\mathbf{v})=S_e,
\label{energyequation}
\end{equation}
\begin{equation}
\frac{\partial\mathbf{B}}{\partial t}+\nabla\cdot (\mathbf{v}\mathbf{B}-\mathbf{B}\mathbf{v})=\mathbf{S}_{B}.
\label{magneticequation}
\end{equation}
$S_{\rho}$, $\mathbf{S}_{\rho v}$, $S_e$, and $\mathbf{S}_{B}$ are all the additional source and sink terms where additional physics can be included. The total pressure $p_{\text{total}}=(\gamma-1)(e-\rho v^2/2-B^2/2)+B^2/2$ is the sum of the gas pressure and magnetic pressure. For the GEM challenge problem Hall-MHD is used which will be described in Sec. 4. The HLLC scheme~\cite{ToroSpruce1994} is used for the spatial discretization while a three-step Runge-Kutta scheme is utilized for the temporal discretization. The simulations are performed in a 2D cartesian geometry with periodic boundary conditions on all four boundaries.

\begin{figure}[h]
\centering
\includegraphics[width=0.9\textwidth]{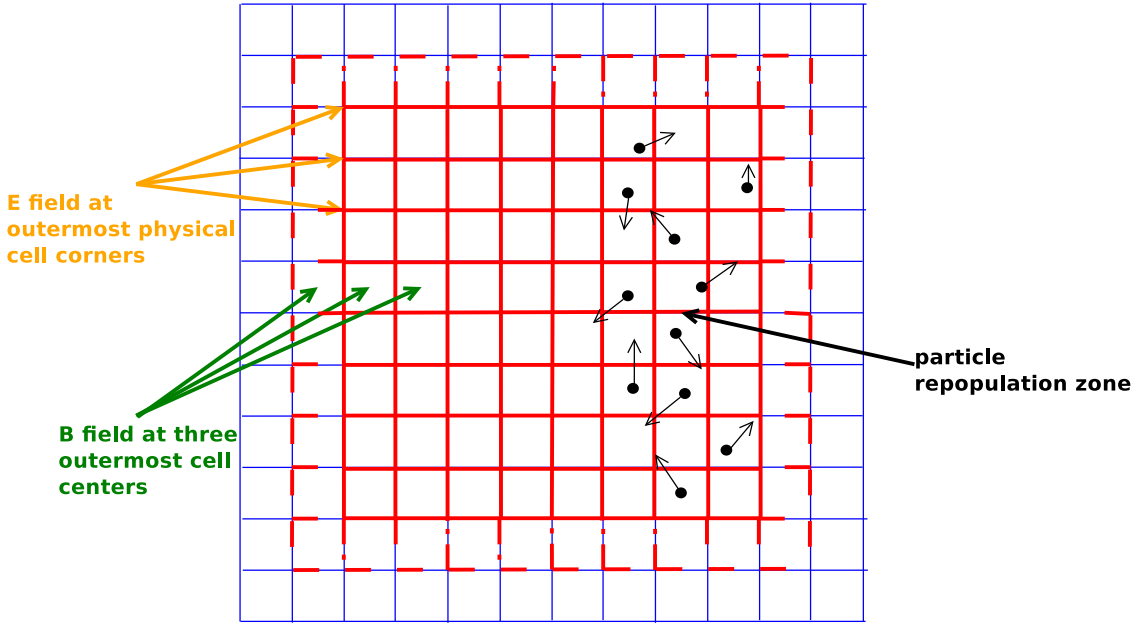}
\caption{A diagrammatic representation of the grid setup and coupling boundary conditions. The thin blue lines represent the MHD grid with the PIC grid lying exactly on top of it in thick red lines. There is a ghost layer of one cell thickness around the PIC grid shown in dashed red lines. The PIC electric field is set on the PIC boundary at the cell corners of the outermost active cells, while the magnetic field is set at the cell centers of the ghost layer and the two outermost active cells. The three outermost active cells also form a particle repopulation zone where particles are deleted and re-injected at every time-step as described in the text.}
\label{grid}
\end{figure}

The MPI-AMRVAC simulation is started with user-specified initial conditions. At a user-specified time a user-specified region in the simulation is selected to be also simulated with PIC. Currently this time and region has to be pre-selected by the user. The PIC simulation can also be started along with the MHD simulation. The PIC region has to be made up of complete grid blocks which form a rectangular region and these grid-blocks have to be at the same level of refinement. At this starting time, the primitive variables of MHD (density, pressure, velocity, and magnetic fields) are gathered and passed to iPIC3D. Using this information the iPIC3D simulation is initialized. The iPIC3D grid is made to exactly superimpose on the MPI-AMRVAC grid, such that cell-centers of iPIC3D grid lie on cell-centers of MPI-AMRVAC grid, as shown in Fig.~\ref{grid}. { In Ref.~\cite{DaldorffToth2014} the MHD grid cell-centers were used as the PIC grid cell-corners, whereas Ref.~\cite{TothJia2016} generalized this to any placement of the PIC grid with respect to MHD grid}. The MHD primitive variables are represented on the cell-centers, whereas several of these variables are required at the cell-corners by iPIC3D. Therefore, the neighboring cell-center average is used to pass the values of these primitive variables at the cell-corners to iPIC3D.

It is important to mention how the units are handled when passing data from one code to another, since the two codes can use different units. { MHD equations do not have any intrinsic units and MPI-AMRVAC simulations are therefore run in dimensionless units. The user can attribute a length scale, reference velocity and a reference density, thereby setting the time and mass units. The dimensionless MHD simulation data is normalized with these normalization factors such that the information passed is in iPIC3D units. For this, the} length is normalized in units of ion-skin depth $d_i$, and the velocity is normalized to the speed of light, $c$. Therefore, the time normalization is inverse ion plasma frequency $\omega_{p,i}^{-1}=d_i/c$. The mass density in MPI-AMRVAC is normalized to some background density $\rho_0$, so the background density is unity in code units. This mass density in MPI-AMRVAC is used to derive the particle number density in iPIC3D assuming the mass of ions ($m_i$) to be unity and a user defined ion to electron mass ratio. The pressure in MPI-AMRVAC and iPIC3D is normalized to $m_{i}c^{2}/d_{i}^{3}$. The magnetic field is normalized to $c\sqrt{4\pi\rho_{0}}$ if its value is known in C.G.S. units or to $c\sqrt{\mu_0\rho_0}$ if its value is known in S.I. units, where $\mu_0$ is the permeability of free space. This leads to the magnetic field in code units being equal to $v_A/c$, where $v_A$ is the Alfv{\'e}n speed { for the background mass density $\rho_0$}. 

iPIC3D uses the MHD grid and its primitive variables to initialize its simulation. The initial magnetic field at the cell corners and cell centers is directly taken from the MPI-AMRVAC data, { using bilinear interpolation in 2D to obtain the values at any point from the neighboring cell-center values of MHD}. In iPIC3D the magnetic field is advanced using the induction equation which adds the curl of electric field to the magnetic field to advance it. As a result the $\nabla\cdot\mathbf{B}=0$ condition of magnetic field is maintained by default. { MPI-AMRVAC has a variety of divergence control methods to keep $\nabla\cdot\mathbf{B}=0$. For the wave simulations shown below, we use the diffusive approach of~\cite{KeppensNool2003},\cite{Marder1987}, whereas for the reconnection simulations we follow a hyperbolic divergence cleaning method (glm3) as in ~\cite{PorthXia2014}\cite{DednerKemm2002}}. { Electric field is calculated in MPI-AMRVAC by the Hall-MHD Ohm's law, $\mathbf{E}=-\mathbf{v}\times\mathbf{B}+\eta\mathbf{J}+(\eta_h/\rho)(\mathbf{J}\times\mathbf{B})$, where $\eta$ is the resistivity and $\eta_h$ is the Hall parameter (details in Sec.~\ref{whistlerwaves}). This is the general form of the electric field used to initialize the electric field in the PIC simulation on the cell corners. In the simulations where we have used the MHD module, we set the resistivity ($\eta$) and Hall parameter ($\eta_h$) equal to zero. In simulations with Hall-MHD module, the $\eta_h$ parameter is set to some finite value, while $\eta$ is still zero. However, the resistivity can also be used if collisional resistivity is introduced in the PIC simulations in the future.} In iPIC3D, each cell starts with a certain number of particles per cell per species. We limit ourselves to a two-species plasma with equal and opposite charge on the two species. The charge is normalized to proton charge ($q_i$). The mass of the positive charge (ion mass $m_i$) is taken as unity and the ratio of ion mass to electron mass ($m_e$) is a user-defined parameter. {The particles are initialized in a way similar to Eqs. 15-24 of Ref.~\cite{DaldorffToth2014}, except that since MPI-AMRVAC gives an isotropic pressure, the velocity distribution is isotropic}. The number density of ions ($n_i$) is equal to the number density of electrons ($n_e$) by charge neutrality and is set according to 
\begin{equation}
n\equiv n_i=n_e=\frac{\rho_{\mathrm{MHD}}}{m_i+m_e},
\label{setdensity}
\end{equation}
where $\rho_{\mathrm{MHD}}$ is the density value received from MPI-AMRVAC.
 
The particles are initialized with a velocity ($\mathbf{v}$) given by
\begin{equation}
\mathbf{v}=\mathbf{u}_0+\mathbf{u}_{th}
\label{u0+vth}
\end{equation} 
Here $\mathbf{u}_0$ is a drift velocity which is derived from the flow and current density values supplied by MPI-AMRVAC. The current density, $\mathbf{J}$, is simply calculated by $\nabla\times\mathbf{B}$ in MPI-AMRVAC and also passed to iPIC3D. The drift velocity of species s, $\mathbf{u}_{0,s}$ is,
\begin{equation}
\mathbf{u}_{0,s} = \frac{[1+({m_{s'}}/{m_{s}})][({q_{s'}}/{m_{s'}})\mathbf{u}_{\mathrm{MHD}}-({\mathbf{J}}/{\rho_{\mathrm{MHD}}})]}{[({q_{s'}}/{m_{s'}})-({q_{s}}/{m_{s}})]}
\label{driftspeed}
\end{equation}
Here $s'$ is the other species with charge $q_{s'}$ and mass $m_{s'}$, $\mathbf{u}_{\mathrm{MHD}}$ is the fluid flow velocity from MPI-AMRVAC. Eq.~\ref{driftspeed} is derived by expressing the MHD flow and current density in terms of contributions from the two-species velocities and then solving for the two-species velocities. The particles are also assigned a random thermal velocity $\mathbf{u}_{th}$. This is done by randomly assigning a thermal speed ($v$) to the particles in each cartesian direction. A uniform distribution of random variables $r$ between 0 and 1 is converted into a Maxwellian distribution of speeds $v_{d}$ by~\cite{BoxMuller1958},
\begin{equation}
v_d=v_{th}\sqrt{-2\ln(1-\kappa r)}.
\label{distributiongeneration}
\end{equation}
Here $\kappa$ is a number very close to, but slightly smaller than, unity to avoid numerical problems with the logarithm function. $v_d$ is further multiplied by $\cos(\theta)$($\sin(\theta)$) to get the velocity $v$ in the cartesian $x$($y$) directions, where $\theta$ is a uniformly distributed random variable between 0 and $2\pi$. Similar procedure is used in the third direction, and this can also handle anisotropic distribution in the two directions. However, in this work we are assuming an isotropic distribution, using a scalar pressure from MHD. This produces the following speed distribution function in each cartesian direction,
\begin{equation}
f(v)=\bigg(\frac{2v}{v_{th}^2}\bigg)\exp\bigg(\frac{-v^2}{v_{th}^{2}}\bigg).
\label{velocitydistribution}
\end{equation}
Here $v_{th}$ is the thermal velocity which is derived from the MHD pressure and density. The species dependent thermal velocity is
\begin{equation}
v_{th,s} = \sqrt{\frac{p_{s}}{\rho_{\mathrm{MHD}}}\bigg(1+\frac{m_{s'}}{m_{s}}\bigg)}.
\end{equation}
$p_s$ is the pressure of species $s$. This is derived from MPI-AMRVAC, however MHD only provides a single fluid pressure and we have the freedom to distribute this pressure between the two particle species depending on what temperature ratio we want between the two species. The pressure is divided between ions and electrons as 
\begin{eqnarray}
p_e=\zeta p, \\
p_i=(1-\zeta)p,
\end{eqnarray}
where $\zeta$ is the fraction of pressure supported by electrons. Thus, assuming quasineutrality, the temperature ratio is $T_i/T_e=(1-\zeta)/\zeta$.




The time-dependent boundary conditions to the PIC simulation are provided from MHD as shown in Fig.~\ref{grid}. At every time step, MPI-AMRVAC supplies the values of its physical quantities at the boundary of the PIC region as described in the time-stepping method (Sec. 2.3). { The PIC domain contains one layer of ghost cells around it, shown by the dashed red lines in Fig.~\ref{grid}. All cells inside of this layer are called active cells, as they actively evolve the fields and particles. The ghost cell contains the information of the magnetic field at the cell-centers and also particles that leave the active domain. The magnetic field at the ghost cell center is set from MHD values. The magnetic field at the cell centers is used in solving the second order Amp{\`e}re's law for the electric field at the cell nodes. In the field solver the electric field on the outermost active nodes of the PIC domain (indicated by yellow arrows in Fig.~\ref{grid}) is a boundary condition and is set from the MHD value~\cite{RicciLapenta2002}. After this the magnetic field is advanced by the induction equation. The magnetic field on the cell-centers of the ghost layer is already specified as the MHD value. In addition the magnetic fields at the two outermost active cell centers (indicated by the green arrows, including the ghost cell center, in Fig.~\ref{grid}) are also now changed to the MHD values. This was seen to improve the solution at the boundary. The particles which enter the PIC ghost layer can be treated according the different boundary conditions, like reflection or exit or reemission with some random thermal velocity. In this case, we remove the particles that enter the ghost layer. Instead, the three outermost active cells (indicated by black arrows in Fig.~\ref{grid}) form a particle repopulation zone in which all the particles are deleted at every time step and new particles introduced with a density $n$ and velocity $\mathbf{v}$ as shown in Eqs.~\ref{setdensity}-\ref{u0+vth} and their description. This serves to recreate the MHD solution in the repopulation zone in the velocity space at every time step.}

\subsection{PIC to MHD coupling}
The other half of two-way coupling is PIC to MHD coupling. In this the PIC solution is used to provide feedback to the MHD simulation. This is different from one-way coupling and coupling with particles where MHD fields are used to evolve particles~\cite{RipperdaPorth2017}, but the particle moments are not used to update the MHD simulation. To obtain fluid quantities for feedback, moments of the velocity distribution function have to be taken. iPIC3D calculates the fluid moments from the particle distribution as described in Ref.~\cite{MarkidisLapenta2010}. This generates the primitive variables needed by MHD. After every time step these quantities are passed back to the MHD code. MPI-AMRVAC advances the conservative variables as in Eqs.~\ref{densityequation}-\ref{magneticequation}. Therefore the iPIC3D moments are converted into conservative variables and these are then used to update the MHD solution in the PIC domain. However, we observed that directly replacing the old MHD variables by the PIC moments in the entire PIC region creates some distortions at the MHD and PIC interface (see Sec.~\ref{whistlerwaves}). This sometimes also leads to $\nabla\cdot\mathbf{B}=0$ problems. Therefore, the transition from PIC zone to MHD zone should be made smoothly. This is done by taking a weighted average of the MHD state and the PIC moments, { similar to Ref.~\cite{UsamiOhtani2008}}. A conservative variable $\psi_{\mathrm{MHD}}$ of the MHD solution in the PIC zone is replaced by $\hat{\psi}$ as follows
\begin{equation}
\hat{\psi}=(1-w)\psi_{\mathrm{MHD}}+w\psi_{\mathrm{PIC}}
\label{weightedaverage}
\end{equation}
$\psi_{\mathrm{MHD}}$ is one of the conservative MHD variables like mass density ($\rho$), momentum density ($\rho\mathbf{v}$), magnetic field $\mathbf{B}$, or energy density $e$. $\psi_{\mathrm{PIC}}$ is the same conservative variable derived from the PIC moments. $w$ is the weight given to the PIC solution and $1-w$ is the weight given to the MHD solution. This weight function $w$ is chosen such that the MHD solution gets unity weight at the boundary of PIC zone, and zero weight to the PIC solution. Then this function rapidly rises close to unity within a transition layer which surrounds the PIC zone boundary, inside the PIC zone. Interior of this transition layer, the weight is very close to unity and almost uniform, meaning that the MHD solution is almost entirely overwritten by the PIC solution, with the MHD solution having negligible weight. This is different from the coupling with BATS-R-US where this transition layer is not used. Various weight functions have been used in our tests and the following was found to work satisfactorily,
\begin{eqnarray}
w=(1-\exp (-(x-x_{1})^2/\delta^2))(1-\exp (-(x-x_{2})^2/\delta^2))\nonumber \\
     \times (1-\exp (-(y-y_{1})^2/\delta^2))(1-\exp (-(y-y_{2})^2/\delta^2))
\label{weightedfunction}
\end{eqnarray}
Here the PIC simulation domain is bounded by $x_1\le x \le x_2$ and $y_1 \le y \le y_2$. We can see that this weight function falls exactly to zero at the boundaries. Parameter $\delta$ sets the width of the transition layer. Outside the transition layer, i.e., $|x-x_{1,2}|>\delta$ and $|y-y_{1,2}|>\delta$, this function is close to unity. iPIC3D is capable of handling large PIC domains, of the order of several hundred ion skin-depths ($d_i$). In this work, we have PIC zones of the order of several tens $d_i$'s, whereas the transition layer widths $\delta$ are around couple of $d_i$, indicating that only a thin transition layer is required for smooth PIC to MHD coupling.



\subsection{Time stepping method}
\begin{figure}[h]
\centering
\includegraphics[width=1.0\textwidth]{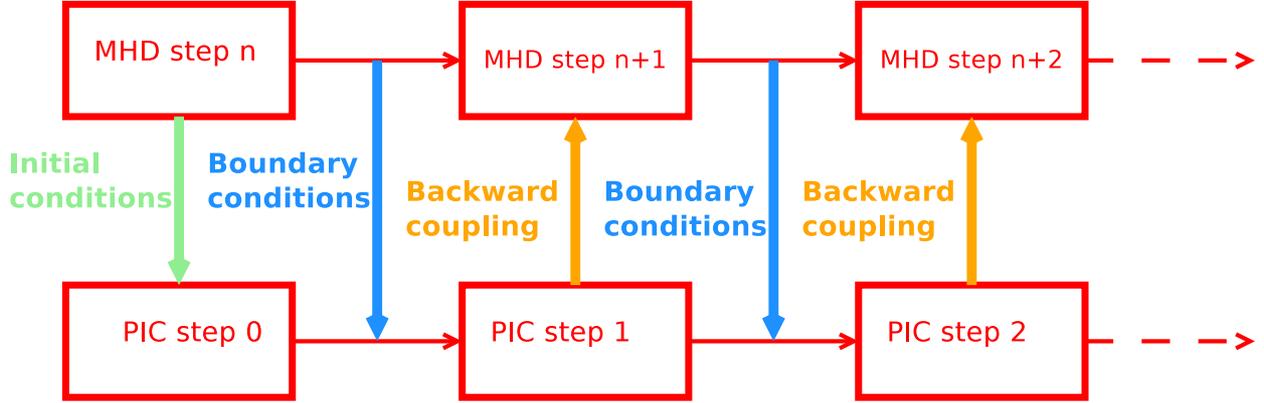}
\caption{Cartoon of the time stepping method. The PIC simulation can be started at some MHD time step $n$. At this stage, initial conditions are passed to PIC step 0. Then MHD goes to step $n+1$, and then the time averaged boundary conditions are provided for the PIC time advance step. After PIC time step 1 the PIC solution is fed back to MHD to update the MHD solution at time $n+1$. This process then repeats for subsequent steps.}
\label{timestep}
\end{figure}
Now we describe how these different parts couple together and advance in time. This exchange of information takes place between MHD and PIC at every time step. The PIC time step is fixed, whereas the MHD simulation has a variable time step depending on the CFL condition. The PIC time step is fixed to be smaller than the MHD time step { which is constrained to be under the CFL limit. The MHD time step is lowered below the CFL limit such that it is an integer multiple of the PIC time step.} In all simulations presented here, the setup has been such that the PIC and MHD time step are equal. However, { when the MHD time step is an integer multiple of the PIC time step, then multiple PIC steps fit within one MHD step.} The time stepping scheme is shown in Fig.~\ref{timestep}. The MHD simulation is started and at some user-specified MHD time step $n$ the PIC simulation is started. At this time initial conditions are passed to iPIC3D to initialize the PIC simulation. Then the MHD time step is taken to advance the MHD state to step $n+1$. The values of the MHD variables at the PIC domain boundaries are averaged between steps $n$ and $n+1$ and passed as boundary conditions to iPIC3D. These are the boundary conditions utilized to advance the PIC state to step 1 as discussed in MHD to PIC coupling (Sec.~\ref{MHDtoPIC}). This will be second-order accurate in time, { compared to first order coupling when the MHD state at step $n$ provides the boundary conditions for the PIC simulation to advance to MHD step $n+1$, as used in the coupling with BATS-R-US. In case the MHD time step is made up of multiple PIC steps, then the boundary conditions provided to each individual PIC step will be linearly interpolated between the MHD states at steps $n$ and $n+1$. In the simulations shown here, where 1 MHD time step was equal to 1 PIC time step, we have tried changing the coupling to first order, and it produces very similar results to the second-order in time coupling. This is because iPIC3D itself is accurate to only first-order in time. However, in cases when the MHD time step grows very long or when it consists of multiple PIC time steps, then such a time interpolation scheme could produce better coupling.} After the PIC steps have completed, moments in iPIC3D are calculated and passed back to MPI-AMRVAC to implement the PIC to MHD coupling where the MHD solution in the PIC domain at step $n+1$ is overwritten by the weighted average of Eq.~\ref{weightedaverage}. After this the MHD simulation is advanced to step $n+2$ and the whole process is repeated. Data is output at regular intervals in both MHD and PIC simulations for analysis. 


\subsection{Parallelization}
{ MPI-AMRVAC is a block-adaptive code in which the computing unit is a block of grid-cells. These blocks are distributed among computing processors according to the Morton order curve}, which tries to place adjacent grid blocks on the same processor to minimize communication costs in an adaptively refined mesh. iPIC3D uses a straightforward cartesian domain decomposition to distribute cells on various processors. Moreover, the PIC domain is typically only a small subsection of the entire MHD domain. {If we distribute the PIC simulation only over those processors which handle the PIC domain in the MHD simulation, we will end up using only a small fraction of the total available number of processors for advancing the PIC simulation, while the rest of the processors wait for the updated PIC result. We could also divide the MHD and PIC simulations on two different sets of processors, but that would mean that the MHD set of processors sits idle while PIC computations are being made, and vice-versa. Instead, in this coupling the PIC simulation is made a part of the MHD simulation by building it as a library which can run concurrently with MPI-AMRVAC on the same processor.} Therefore, we can distribute the PIC domain over all (or the maximum number possible, depending on how best the domain decomposition can fit on the number of processors) of the available processors which also share the entire MHD domain. This means that each processor is handling a part of the MHD simulation as well as part of the PIC simulation. This way, no processor sits idle, it alternately advances the MHD simulation followed by PIC followed by MHD and so on, { thereby utilizing most of the processors efficiently}.  

{ The drawback of this approach is more communication cost between processors, since the distribution of MHD and PIC domains over processors varies significantly. A single processor is typically processing very different parts of the physical domain. At every time step, data has to be communicated between MPI-AMRVAC and iPIC3D over all these processors. This is done through a master MHD processor where all required MHD data is gathered and then redistributed to the PIC routines as needed, and similarly through a master PIC processor. The exchange of information between these MHD and PIC master processors is through MPI. This is certainly not optimal if there are a large number of processors, but for the 2D simulations shown here, it takes only a small fraction of the computing time.} This method will be optimized in further revisions, making the simulation cost effective and ready for large applications. Below we focus on testing the physics of this coupling methodology with regards to energy and momentum conservation and wave propagation. 



\section{Physics tests}

\subsection{Energy and momentum conservation and stability}

\begin{figure}[h]
\centering
\includegraphics[width=0.9\textwidth]{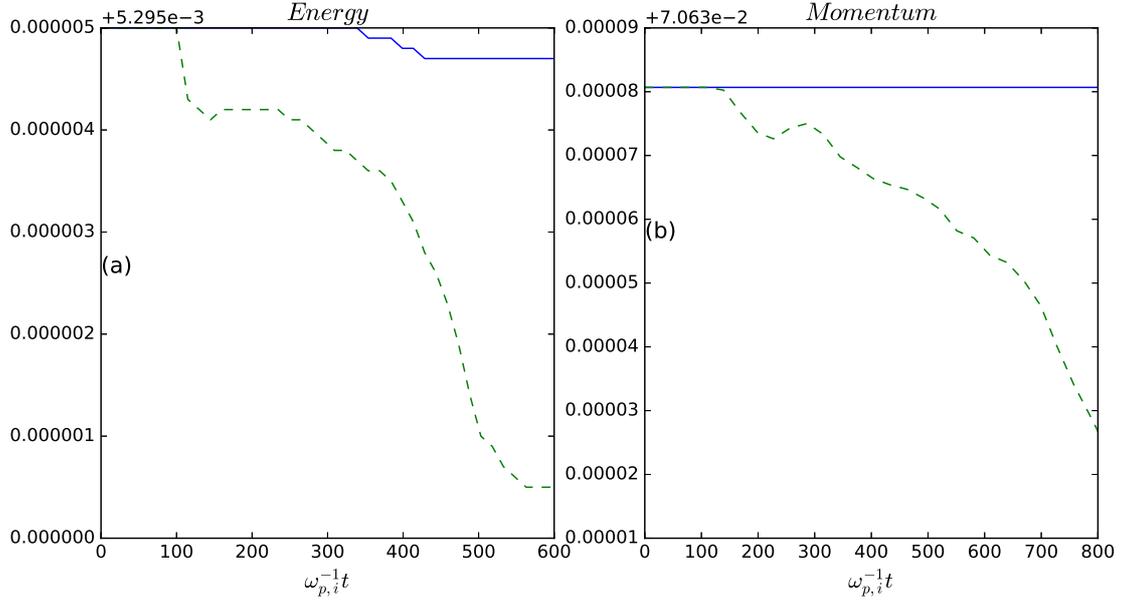}
\caption{(a) shows the time evolution of energy in a setup with a force-free steady-state current sheet. The solid line is a pure MHD run, the dashed line is with a two-way coupled simulation. The energy dissipates by less than 0.1\% in the two-way coupled simulation. (b) shows the total momentum evolution in a simulation with a constant flow. Here also the two-way coupled simulation (dashed line) shows a decay in momentum of less than 0.1\%.}
\label{conservation}
\end{figure}

We first started with testing of steady-state equilibrium setups to verify conservation of energy and momentum and also the stability of this coupling. To test the coupling of electromagnetic fields, we setup a force-free current sheet. The sheet has $B_x=B_0\tanh((y-L_y/2)/\lambda)$ and $B_z=B_0\sech((y-L_y/2)/\lambda)$. The simulation box size is $L_x=100d_i$ and $L_y=80d_i$. The PIC domain is of size $(40d_i\times40d_i)$ and it contains the current sheet. The resolution in MHD is $250\times 200$ cells and $100\times 100$ cells in PIC. Same cell sizes are used in MPI-AMRVAC and iPIC3D. There are 1000 particles per PIC cell, the ion-electron mass ratio is 20, and ion to electron temperature ratio is unity. The magnetic field amplitude is $B_0=0.1$ and sheet thickness is $\lambda=5d_i$. The current sheet remains stable throughout the simulation and the interface between MHD and PIC regions does not develop any errors, showing good coupling of the electromagnetic fields. There are small fluctuations due to the PIC particle noise, these can be reduced by increasing the number of particles. Fig.~\ref{conservation}(a) shows the evolution of total energy in this simulation compared to a pure MHD simulation. The energy is not perfectly conserved in the pure MHD simulation because there is a conducting boundary condition at the $y$ boundary. In the two-way coupled simulation, we see that the energy conservation is very good, to within 0.1\%. 

We also simulate a steady flow in the simulation box whose velocities are $v_x=0.03c$, $v_y=0.04c$, and $v_z=0.05c$. The other setup parameters of the simulation box are same as above. The flow tests the particle boundary conditions, especially the particle repopulation zone implementation, in all the three directions. The streamlines in MHD and PIC regions are perfectly aligned, with the magnitude of the velocities also perfectly matched. MPI-AMRVAC has no fluctuations whereas iPIC3D develops some fluctuations with r.m.s. values less than 10\% of the mean velocity. Fig.~\ref{conservation}(b) shows the total momentum in the simulation domain. We see that the total momentum in the two-way coupled simulation is conserved to within 0.1\% of the pure MHD simulation. This is true for the three momentum components individually also, thereby showing excellent coupling between the two codes for particles also.

\subsection{Circularly polarized Alfv{\'e}n waves}
\begin{figure}[h]
\centering
\includegraphics[width=1.0\textwidth]{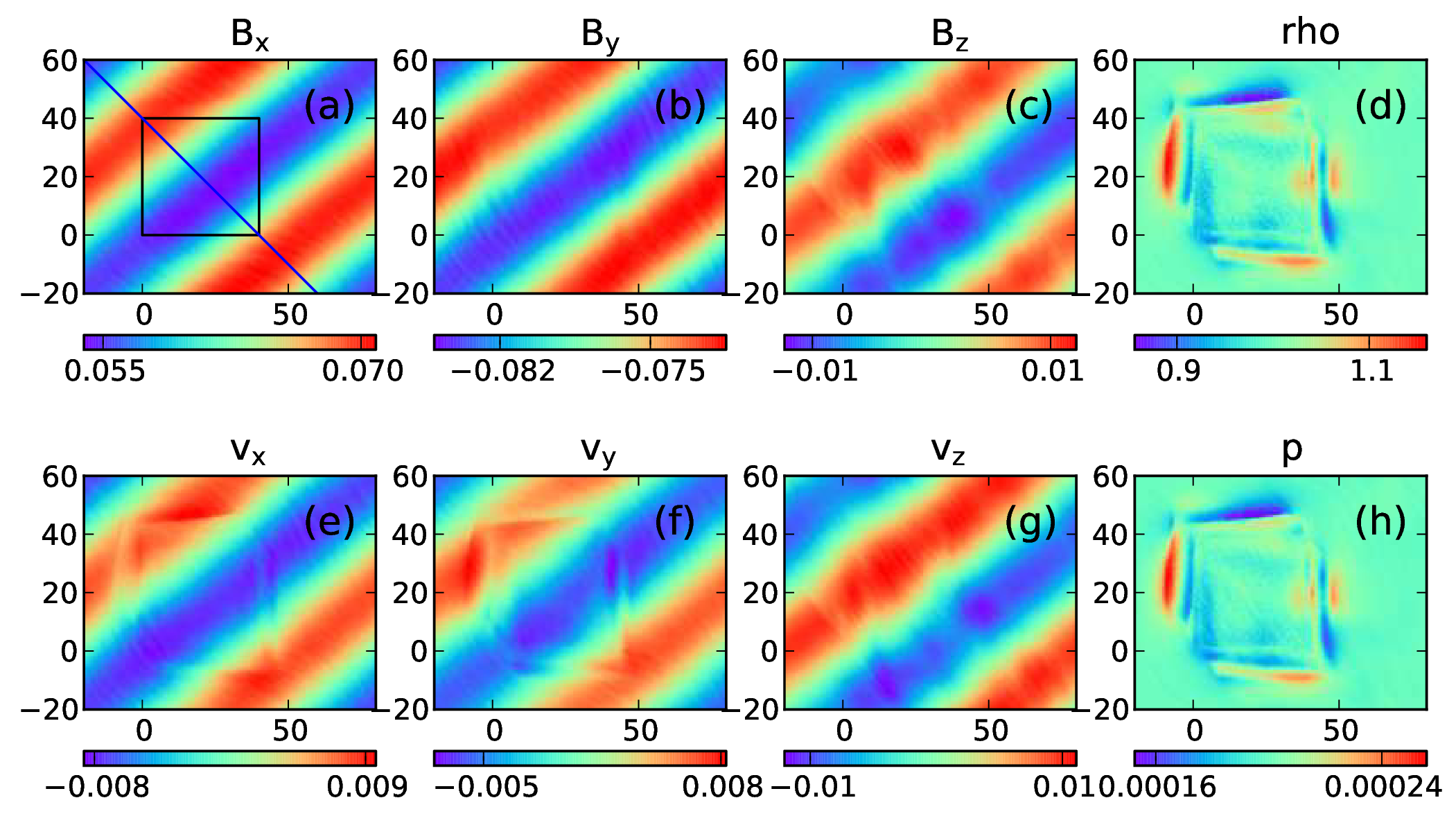}
\caption{The 8 primitive variables of MHD shown for the Alfv{\'e}n wave after simulating $1.12$ wave period with the two-way coupling. (a) $B_x$, (b) $B_y$, (c) $B_z$, (d) $\rho$, (e) $v_x$, (f) $v_y$, (g) $v_z$, and (h) $p$. The PIC region is shown by the black box in the $B_x$ figure. The box length is in terms of ion skin depth $d_i$.}
\label{alfven_color}
\end{figure}
To test this code coupling, we start with simulating the well-known plasma MHD waves. We take a circular Alfv{\'e}n wave with the following form:
\begin{eqnarray}
\rho = \rho_0 \\
p=p_0 \\
B_x=B_0 \\
B_y=\delta B\cos(kx) \\
B_z=\delta B\sin(kx) \\
v_x = 0 \\
v_y = \delta u \cos(kx) \\
v_z = \delta u \sin(kx)
\end{eqnarray}
The simulation is carried out in a box with $L_x=100d_i$ and $L_y=80d_i$, which stretches across $-20\le x \le 80$ and $-20\le y \le 60$. The resolution is $500\times 400$ cells giving $\Delta x = \Delta y =0.2d_i$. The PIC domain is of $40d_i\times 40d_i$ size lying between $0\le x \le 40$ and $0\le y \le 40$ as shown by the black box in Fig.~\ref{alfven_color}. The transition width $\delta$ specified in Eq.~\ref{weightedfunction} is set to $2d_i$. The PIC simulation has 1000 particles per species per cell. The ion-to-electron mass ratio is taken as 20 and also $T_i=T_e$, so the pressure is distributed equally between ions and electrons. The simulation box is rotated by an angle $\theta=\arctan (L_x/L_y)$ w.r.t. the $x$-axis to test both the dimensions. The wavenumber is taken as $k=\sqrt{(2\pi/L_x)^2+(2\pi/L_y)^2}$ so that the wave satisfies the periodic boundary conditions in the tilted simulation box. The wave period is $\omega=kv_A$, where $v_A$ is the standard Alfv{\'e}n velocity, which in normalized units is equal to $B_0$. $B_0$ is taken as 0.1, $\rho_0=1$, $p_0=0.0002$, $\delta B=0.01$, $\delta u =0.01$. The wave time period is $\tau_A=2\pi/(kv_A)$, which is 624.7 in units of inverse ion plasma frequency ($\omega_{p,i}^{-1}$). The time step is taken to be $\Delta t=0.5\omega_{p,i}^{-1}$. After the start of the MHD simulation, at $t=0$, we initiate the PIC simulation at $t=100$. We show the state of the solution in the two-way coupled simulation at time $t=800$, i.e. after a coupling time of $1.12\tau_A$, in Fig.~\ref{alfven_color}. The magnetic field components show a reasonable coupling without developing significant distortion. However, the velocities show significant distortion generated, mainly at the coupling boundaries. The pressure and density should stay uniform for a circular Alfv{\'e}n wave, but we see perturbations developing along the coupling boundary with significant amplitude of almost 25\% of the background.

\begin{figure}[h]
\centering
\includegraphics[width=1.0\textwidth]{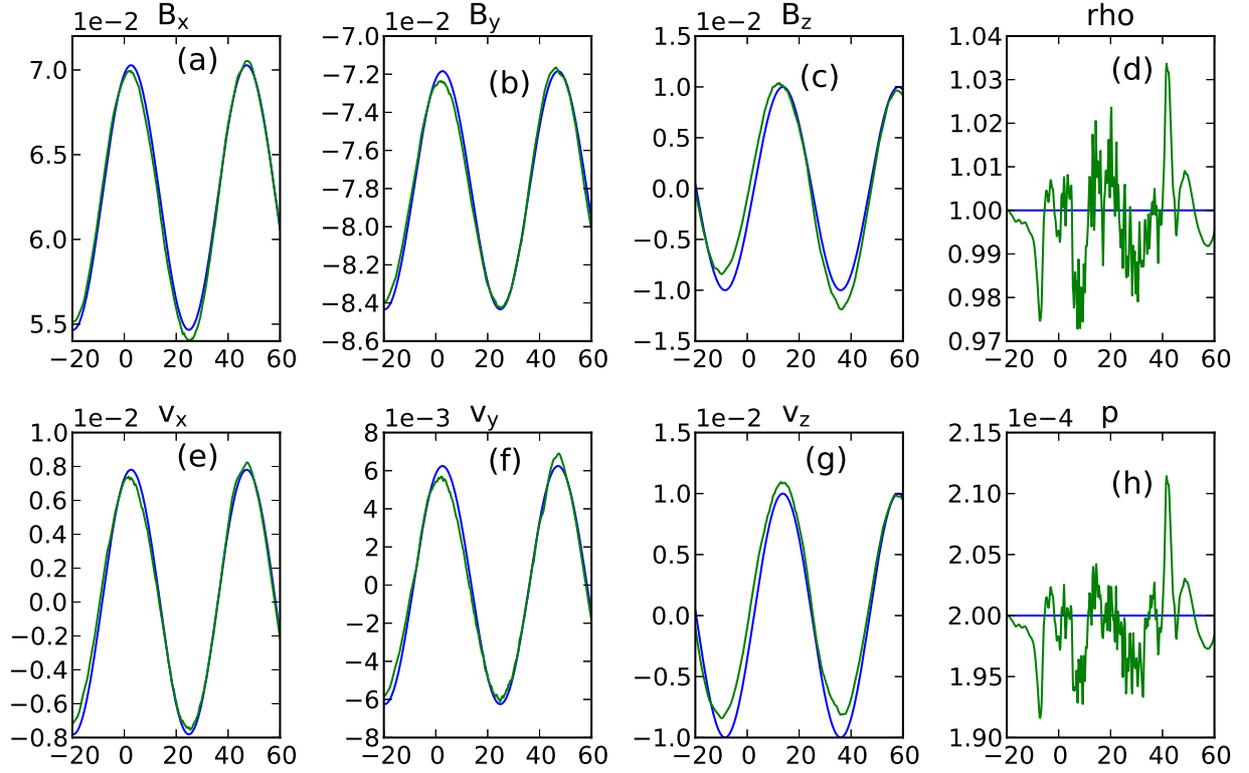}
\caption{A trace of the various quantities through the diagonal line shown in Fig.~\ref{alfven_color}. The $x$-coordinate in the plots is the $x$-coordinate along the simulation box. The PIC region lies within $x=0$ and $x=40$. The blue line is from the MHD-only simulation with the same setup at the same time, whereas the green line is from two-way coupled simulation.}
\label{alfven_line}
\end{figure}
A line cut is taken through the simulation box along the diagonal line shown in Fig.~\ref{alfven_color}(a). This line passes across the diagonal of the PIC domain. The traces of the 8 primitive MHD variables along this line are shown in Fig.~\ref{alfven_line} as a function of the $x$-coordinate along the line. The PIC domain lies within $0\le x \le 40$. A pure MHD simulation was also done with the same setup, without the coupling. The two lines correspond to an MHD-only simulation and the two-way coupled MHD solution. Here also we see that the magnetic field coupling is satisfactory. Otherwise the coupling is not good, especially for the pressure and density variables. Moreover, the mismatch between the two-way coupled simulation and the MHD-only solution also extends outside the PIC domain. The mismatch is worse if we take a cut along the $y=20$ or $x=20$ line, the vertical and horizontal cuts through the center of the PIC domain. We also observed that the coupling gets worse when a smaller box is taken, increasing the kinetic effects. If we take a box with half the length in $x$ and $y$, the wavelength is also halved, then the coupling becomes worse. This indicates that a good match for Alfv{\'e}n waves between MHD and PIC requires very long wavelengths, as was seen in our previous work~\cite{MakwanaZhdankin2015},\cite{MakwanaLi2016}. In the smaller wavelength case we see that the wave in the PIC domain travels faster than the wave in the MHD domain. This can be due to the fact that dispersive whistler physics starts playing an important role in the PIC simulation at smaller wavelengths. Therefore, we investigate the whistler wave coupling in the next section

\subsection{Whistler waves}\label{whistlerwaves}
\begin{figure}[h]
\centering
\includegraphics[width=1.0\textwidth]{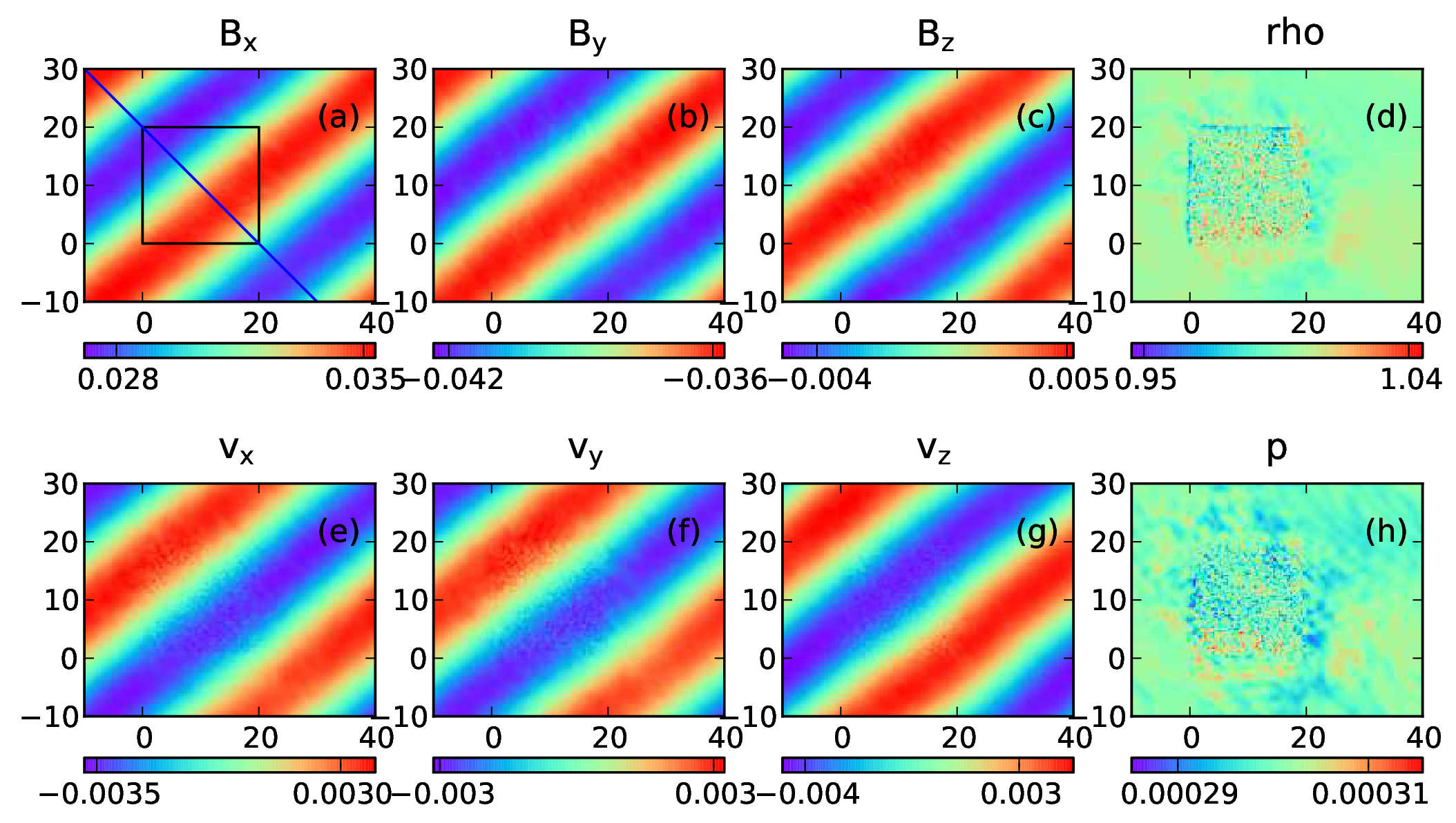}
\caption{The 8 primitive variables of MHD in a two-way coupled simulation of the whistler wave. This is the state of the solution after 1.06 wave period showing very good coupling between MHD and PIC.}
\label{whistler_color}
\end{figure}

The whistler waves are a solution of the Hall-MHD equations. MPI-AMRVAC can solve for the Hall MHD equations by adding the Hall term to the induction equation, Eq.~\ref{magneticequation}, as follows:
\begin{equation}
\frac{\partial \mathbf{B}}{\partial t}+\nabla\cdot\bigg[\mathbf{v}\mathbf{B}-\mathbf{B}\mathbf{v}+\frac{\eta_{h}}{\rho}(\mathbf{B}\mathbf{J}-\mathbf{J}\mathbf{B})\bigg]=-\nabla\times\eta\mathbf{J}
\end{equation}
Here $\eta$ is the resistivity and $\eta_h$ is an adjustable parameter which sets the strength of the Hall-term~\cite{PorthXia2014, LeroyKeppens2017}. In normalized units it is equal to $\bar{\eta}_h=V_{A0}/(L_0\omega_{c,i})$. In these simulations, the Alfv{\'e}n velocity is $V_{A0}=0.05$, the reference length $L_0$ is the ion skin-depth which is unity, and the ion gyrofrequency $\omega_{c,i}=q_iB_0/m_i$ is also 0.05. Therefore $\bar{\eta}_h$ should be unity. However, as this term is treated explicitly in MPI-AMRVAC, it can cause the time step becoming smaller than even the PIC requirement. A value of $\bar{\eta}_h=0.6$ was taken keeping the time steps comparable. The resistivity $\eta$ is set to zero. The simulation is carried out in a box of size $L_x=50d_i$ and $L_y=40d_i$ stretching across $-10 \le x \le 40$ and $-10 \le y \le 30$. The resolution is $250 \times 200$ cells, giving $\Delta x = \Delta y = 0.2d_i$. The PIC simulation domain has a size of $(20d_i,20d_i)$ with the same resolution, stretching across $0 \le x \le 20$ and $0 \le y \le 20$. The interfacing layer width $\delta$ in Eq.~\ref{weightedfunction} is set to $2d_i$. The initial waveform of the whistler wave is taken as~\cite{DaldorffToth2014}:
\begin{eqnarray}
\rho = \rho_0 \\
p = p_0 \\
B_x = B_0 \\
B_y = \delta B \cos(kx) \\
B_z = -\delta B \sin(kx) \\
v_x = 0\\
v_y = -\delta u \cos(kx) \\
v_z = \delta u \sin(kx)
\end{eqnarray}
Here we take $B_0=0.05$, $\delta B=0.005$, $\rho_0=1.0$, and $p=0.0003$. The velocity perturbation amplitude, $\delta u$, is calculated by the following formula:
\begin{eqnarray}
\frac{\delta u}{\delta B} = \frac{B_x}{c_w\rho}
\end{eqnarray} 
where the speed of the wave $c_w=(W/2)+\sqrt{(B_x^2/\rho)+(W^2/4)}$, and $W=(m/q_i)(kB_x/\rho)$. This is taken from Ref.~\cite{DaldorffToth2014}. This waveform is very similar to the circular Alfv{\'e}n wave form except for the small correction of $W$, which makes the wave dispersive and makes it faster than the Alfv{\'e}n wave at smaller wavelengths. However, this makes the waves behave much better in the kinetic code as we show. There are 1000 particles per cell per species and the ion-to-electron mass ratio is 20. The ion and electron temperatures are taken equal $T_i=T_e$. As in the circular Alfv{\'e}n wave case, the reference frame is rotated by an angle $\theta=\arctan (L_x/L_y)$ and the wavenumber is selected to be $k=\sqrt{(2\pi/L_x)^2+(2\pi/L_y)^2}$ so that the periodic boundary conditions are maintained. The time period of this wave is $2\pi/(kc_w)=565.015\omega_{p,i}^{-1}$. We begin the PIC coupling at time $t=50\omega_{p,i}^{-1}$. Fig.~\ref{whistler_color} shows the solution in the two-way coupled simulation after time $t=650\omega_{p,i}^{-1}$, i.e. after evolution of 1.06 wave periods. We see that the PIC domain accurately simulates the Hall-whistler wave, with very little noise being generated throughout the domain and at the boundaries. Furthermore, the coupling is good for all the quantities as can be compared to the whistler wave results in the coupling with BATS-R-US (Ref.~\cite{DaldorffToth2014}, Fig. 5). Some particle noise is generated in the pressure and density plots (Figs.~\ref{whistler_color}(d) \& (h)) which we estimate below.

\begin{figure}[h]
\centering
\includegraphics[width=1.0\textwidth]{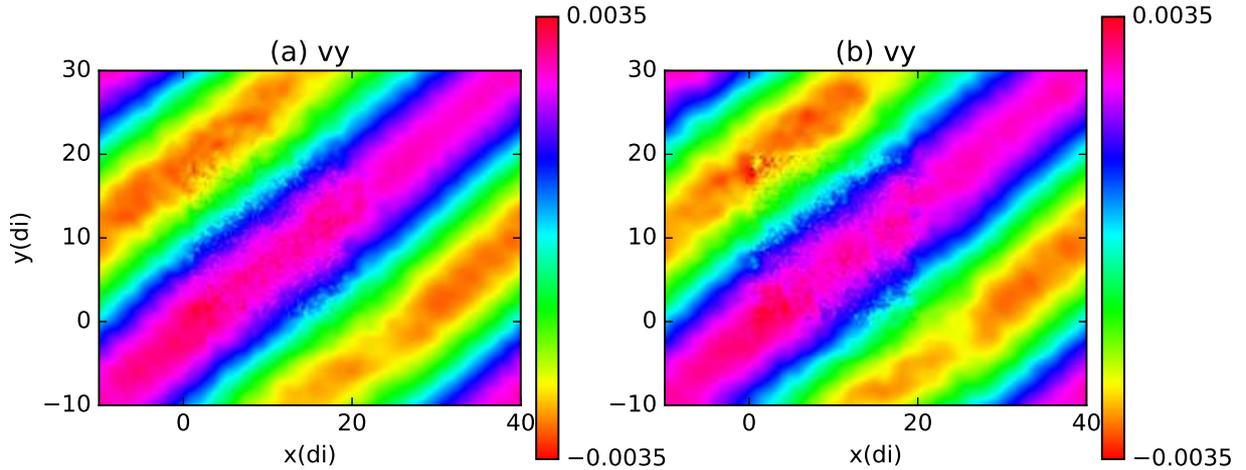}
\caption{The solution in the two-way coupled whistler wave simulation after time $t=960\omega_{p,i}^{-1}$. Both the panels show the velocity in $y$ direction $v_y$. As in Fig.~\ref{whistler_color}, the PIC feedback is in the box $0\le x\le 20$, $0\le y \le 20$.  (a) is the simulation which uses a smoothing layer of thickness $2d_i$ around the MHD-PIC interface when the MHD solution is updated with PIC. In (b) a very thin smoothing layer of width $0.05d_i$ is used, showing that the coupling quality deteriorates.}
\label{whistler_compare_smooth_nonsmooth_BC}
\end{figure}
{ We investigate the effect of the smoothing function $w$ used in the PIC to MHD coupling (Eq.~\ref{weightedaverage}) in Fig.~\ref{whistler_compare_smooth_nonsmooth_BC}. Fig.~\ref{whistler_compare_smooth_nonsmooth_BC}(a) shows the vertical velocity $v_y$ in the same simulation as Fig.~\ref{whistler_color} at a later time $t=960\omega_{p,i}^{-1}$, where a smoothing function of width $\delta=2d_i$ was used. Fig.~\ref{whistler_compare_smooth_nonsmooth_BC}(b) is the same simulation except with a very narrow smoothing function width of $\delta=0.05d_i$. Thus, there is a sharp transition from the MHD to PIC region in this case. We see that the quality of coupling deteriorates with a sharp transition from MHD to PIC. At the MHD-PIC interface, especially at the upper left and lower right corners of the PIC domain in Fig.~\ref{whistler_compare_smooth_nonsmooth_BC}(b) (near $(x,y)=(0,20)$ and $(25,-5)$, in the red color patches) we see the solution is distorted. The PIC simulation develops more fluctuations at the boundaries, and if this is not smoothed out before feeding it back to MHD, then these fluctuations are introduced into the MHD simulation and start advecting throughout the MHD domain. We see these fluctuations also in the density and pressure profiles. The magnetic field components show a better coupling even with a sharp transition layer, but the other velocity components again show some distortion. Thus, we include this smoothing layer in the other simulations also.}

\begin{figure}[h]
\centering
\includegraphics[width=1.0\textwidth]{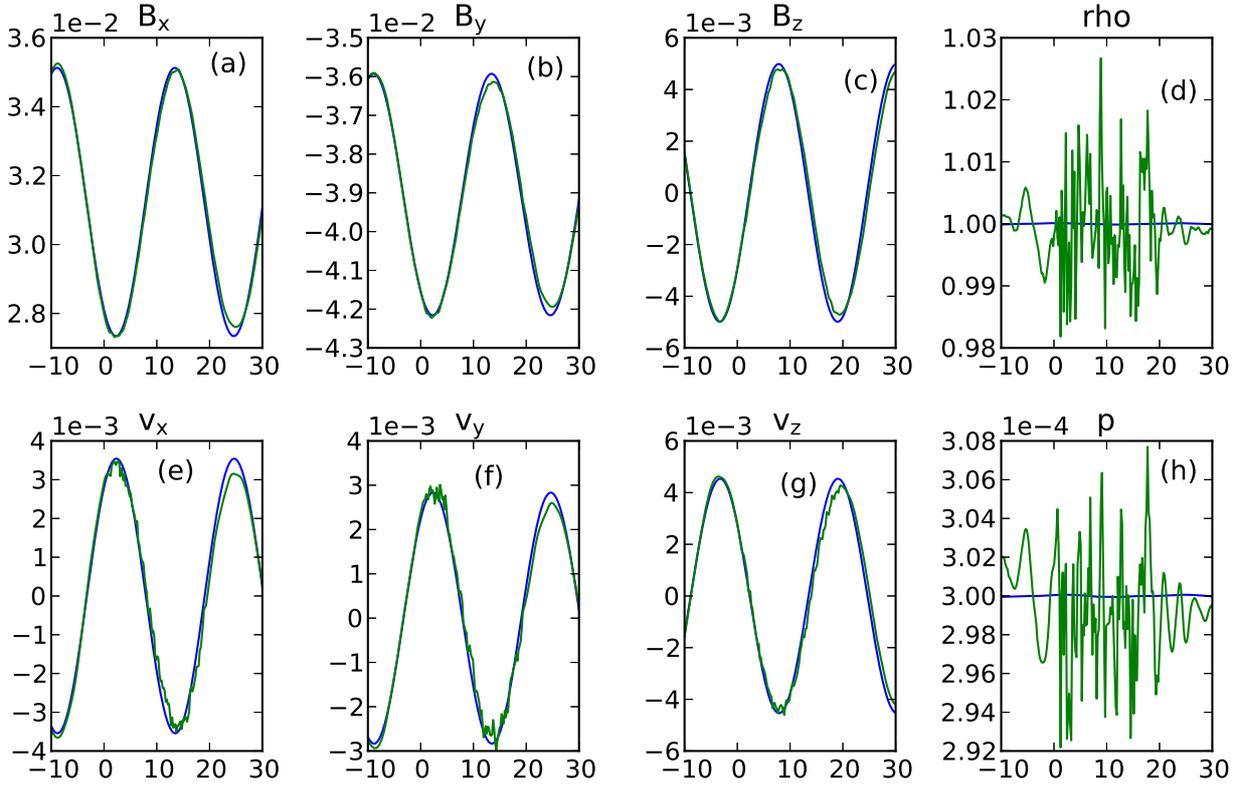}
\caption{Slice through the diagonal line in Fig.~\ref{whistler_color}. The blue curve is from a Hall-MHD simulation without coupling, and the green line is with the two-way coupled code. The two solutions almost lie on top of each other.}
\label{whistler_line}
\end{figure}
As in the Alfv{\'e}n wave case, a trace is taken through the diagonal line shown in Fig.\ref{whistler_color}(a) and the physical quantities along this line are plotted in Fig.~\ref{whistler_line}. We see  very close match with the pure Hall-MHD simulation and the two-way coupled simulation, with both solutions almost lying on top of each other both within and outside the PIC domain, with no visible discontinuity or distortion at the interface. This match remains good for other line-cuts also. The pressure and density fluctuations generated in Figs.~\ref{whistler_line}(d) \& (h) by particle noise is less than 2\% of the background values.

\subsection{Fast magnetosonic wave}
\begin{figure}[h]
\centering
\includegraphics[width=1.0\textwidth]{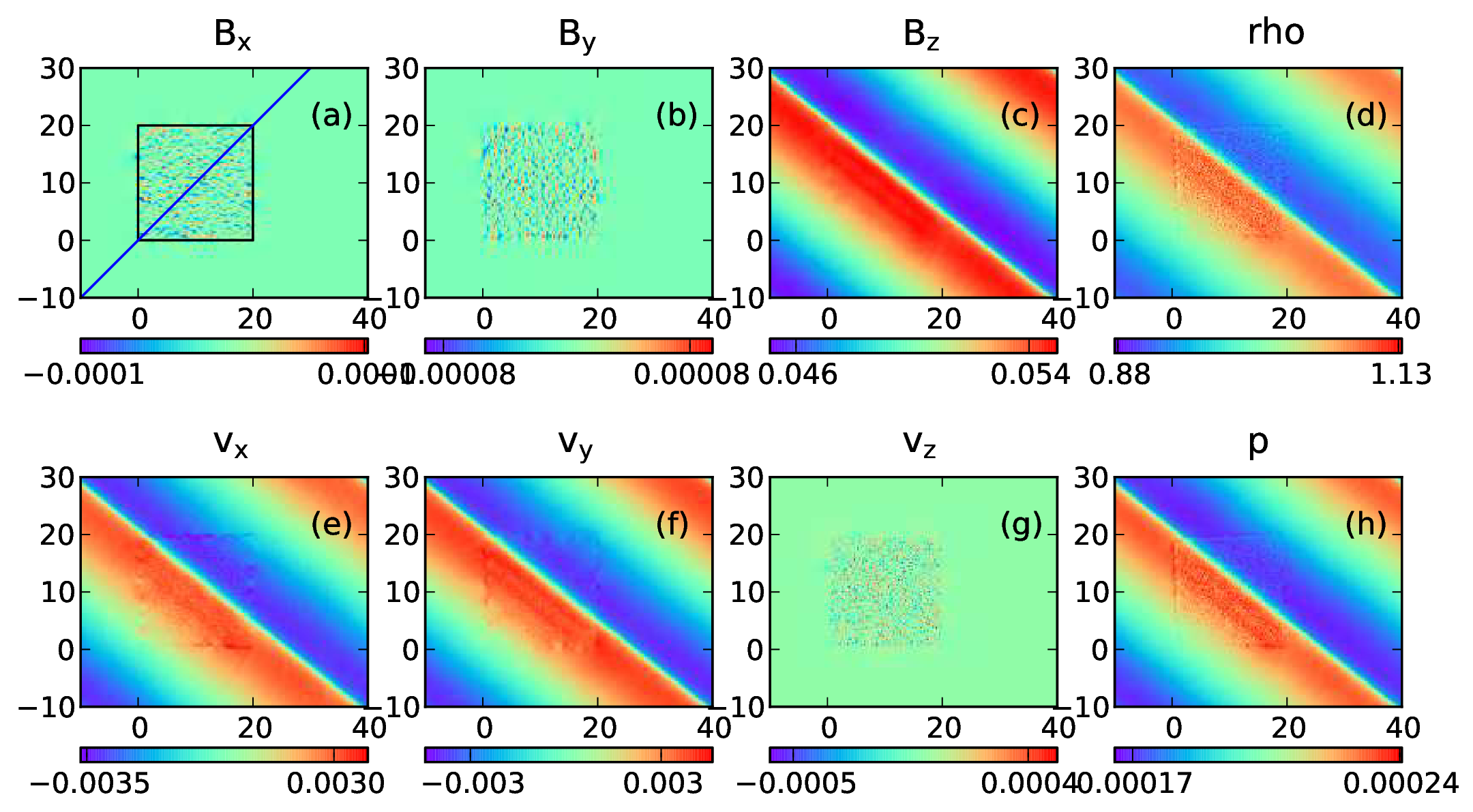}
\caption{Solution of the two-way coupled fast magnetosonic wave after 0.92 wave period.}
\label{magnetosonic_color}
\end{figure}
The pressure and density perturbations are zero in the case of Alfv{\'e}n and whistler waves. To test these we take a look at the fast magnetosonic wave which has perturbations in all the physical MHD quantities of velocity, magnetic field, pressure and density to check their coupling. The initial waveform is 
\begin{eqnarray}
\rho = \rho_0\bigg[1+\frac{k}{\omega}\delta u \sin (k_xx+k_yy)\bigg]\\
p = p_0\bigg[1+\frac{\gamma k}{\omega}\delta u \sin (k_xx+k_yy)\bigg]\\
B_x=0\\
B_y=0\\
B_z=B_0\bigg[1+\frac{k}{\omega}\delta b \sin (k_xx+k_yy)\bigg]\\
v_x = \delta u \frac{k_x}{k}\sin(k_xx+k_yy)\\
v_y = \delta u \frac{k_y}{k}\sin(k_xx+k_yy)\\
v_z = 0
\end{eqnarray}
Here $\rho_0=1$, $p_0=0.0002$, $B_0=0.05$, $\delta u =\delta b =0.005$, $\gamma=5/3$. The wavenumbers are $k_x=2\pi/L_x$ and $k_y=2\pi/L_y$, with $k=\sqrt{k_x^2+k_y^2}$. In this case there is no need to tilt the box as the wavevector itself is aligned along the box diagonal. The wave speed is given by $\sqrt{v_A^2+c_s^2}$, with $v_A=B_0/\sqrt{\rho_0}$ and $c_s=\sqrt{\gamma p_0/\rho_0}$ and its frequency is $\omega = k\sqrt{v_A^2+c_s^2}$. The simulation is performed with the MHD module in MPI-AMRVAC again. The MHD simulation box is $L_x=50d_i$ with $-10\le x \le 40$, $L_y=40d_i$ with $-10\le y \le 30$. The resolution is $(500,400)$ cells, giving $\Delta x = \Delta y = 0.1d_i$. The PIC simulation region is $(20d_i,20d_i)$ with $0\le x \le 20$ and $0\le y \le 20$, with same resolution and 1000 particles per species per cell. The ion-to-electron mass ratio is 20, with the same temperature $T_i=T_e$. The time step for both the codes is $\Delta t = 0.5\omega_{p,i}^{-1}$. The time period of this waveform is $T=586.8\omega_{p,i}^{-1}$. 

\begin{figure}[h]
\centering
\includegraphics[width=1.0\textwidth]{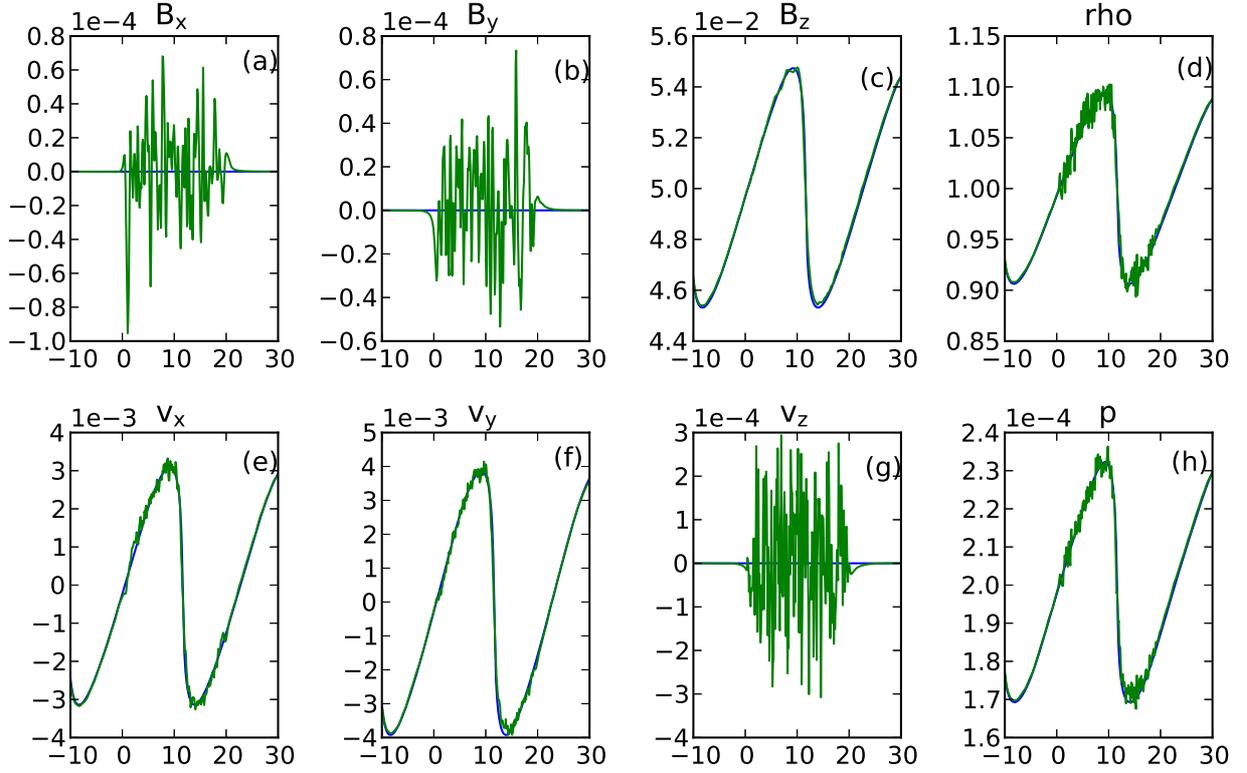}
\caption{Trace of the MHD primitive variables along the line shown in Fig.~\ref{magnetosonic_color}(a) for a fast magnetosonic wave. The pure MHD solution is shown in blue line, with the two-way coupled solution shown in green. This is after 0.92 wave period.}
\label{magnetosonic_line}
\end{figure}
The PIC simulation is started at time $t=60\omega_{p,i}^{-1}$ of the MHD simulation. In Fig.~\ref{magnetosonic_color} we show the various quantities of the fast magnetosonic wave after time $t=600\omega_{p,i}^{-1}$, i.e. after $540\omega_{p,i}^{-1}$ time duration of coupling which is 0.92 wave period. By this time the wave has steepened significantly. However, the coupling looks very good for all the physical quantities, even for the pressure and density perturbation. The steepened wavefront lies within the PIC domain and matches well with the MHD solution outside. It can be compared with the coupling with BATS-R-US (Ref.~\cite{DaldorffToth2014}, Fig. 3). The fast magnetosonic wave is expected to develop a small anisotropy in the pressure tensor, which was accounted for in the coupling with BATS-R-US. Here we have taken an isotropic scalar pressure. However, the results are still very good. Some small distortion is generated at the interface boundary. A better look is shown in Fig.~\ref{magnetosonic_line} which plots the trace through the diagonal line in Fig.~\ref{magnetosonic_color}(a). Here also we see that the two-way coupled solution almost lies on top of the pure MHD solution. It can be compared with Fig.~4 of Ref.~\cite{DaldorffToth2014}. The density and pressure develop noise which is less than 5\% of the background value. The interface between MHD and PIC does not show any noticeable distortion. We have a run a pure kinetic simulation of the magnetosonic wave with the same setup and the entire simulation domain covered by iPIC3D. This also shows very similar behavior as in Figs.~\ref{magnetosonic_color} \& \ref{magnetosonic_line}, but in fact the noise generated is more compared to the two-way coupled simulation. In conclusion, this coupling scheme is working very well for compressible waves also, and also their steepening into shocks. 

{ In this simulation, the area of the PIC domain was $20\%$ of the MHD domain. We also performed a purely PIC simulation with the size of the whole MHD domain, with the same setup, and using periodic boundary conditions. In this case, we get a solution almost the same as the coupled MHD-PIC simulation, showing that the PIC feedback agrees with a global PIC simulation. In an ideal situation, the computing time should be reduced to 20\% if the PIC area is reduced to 20\%. However, there are further costs with communication, modified boundary conditions, etc., and as a result in this case the coupled simulation took 49\% of the computing time compared to the global PIC simulation. The total wall-clock time for this simulation  (up to 0.92 wave period) was 40 minutes on 80 processors of the Flemish Supercomputer Center (VSC) Tier-2 machine. The information exchange between MHD and PIC took $17\%$ of the computing time, the PIC simulation took $79\%$ of the computing time, with the MHD simulation taking the rest. With a better communication strategy, mesh refinement, and smaller PIC domains compared to global domains, the computing cost benefit of this coupling will increase.} Next we apply this coupling to study magnetic reconnection.


\section{Geospace Environmental Modeling (GEM) challenge}
To study magnetic reconnection we implement a setup derived from the Geospace Environmental Modeling (GEM) challenge~\cite{GEM}. This is done with a double current sheet setup, called as a double GEM challenge~\cite{KeppensPorth2013}. The equilibrium is given by
\begin{eqnarray}
\rho = \rho_0 +\sech\bigg(\frac{y-y_1}{\lambda}\bigg)^2 +\sech \bigg(\frac{y-y_2}{\lambda}\bigg)^2 \\
p = \frac{1}{2}B_0^2\rho \\
\mathbf{v}=0 \\
B_x= B_0\bigg\{ -1+\tanh\bigg(\frac{y-y_1}{\lambda}\bigg) +\tanh\bigg(\frac{y_2-y}{\lambda}\bigg)\bigg\}\\
B_y=0\\
B_z=0
\end{eqnarray}
The density $\rho_0=1.0$ and magnetic field $B_0=0.05$. $\lambda=0.5d_i$ is the thickness of the current sheets. The simulation is carried out in a square box of length $L=40d_i$, stretching from $-20\le x \le20$ and $-20 \le y \le 20$. The current sheets are centered at $y_1=-10$ and $y_2=10$. The resolution is $400\times 400$ cells. We have coupled the simulation with the Hall-MHD module in MPI-AMRVAC using a Hall parameter of $\eta_h=1.0$. To this equilibrium a perturbation is added to initiate the reconnection, of the following form:
\begin{eqnarray}
\delta B_x= \psi_{b}k_y\cos[k_xx]\big[\sin(k_y(y-y_1))+2(y-y_1)\cos(k_y(y-y_1))\big]\exp[-k_xx^2-k_y(y-y_1)^2]\nonumber\\
+ \psi_{t}k_y\cos[k_xx]\big[\sin(k_y(y-y_2))+2(y-y_2)\cos(k_y(y-y_2))\big]\exp[-k_xx^2-k_y(y-y_2)^2]\\
\delta B_y = \psi_{b}k_x\cos[k_y(y-y_1)]\big[\sin(k_xx)+2x\cos(k_xx)\big]\exp[-k_xx^2-k_y(y-y_1)^2]\nonumber\\
- \psi_{t}k_x\cos[k_y(y-y_2)]\big[\sin(k_xx)+2x\cos(k_xx)\big]\exp[-k_xx^2-k_y(y-y_2)^2]
\end{eqnarray}
Here $k_x=k_y=2\pi/L$, and $\psi_{t}$ and $\psi_{b}$ are the perturbation amplitudes at the top and bottom current sheets, both are set equal to 0.01 in our simulations. 

\begin{figure}[h]
\centering
\includegraphics[width=1.0\textwidth]{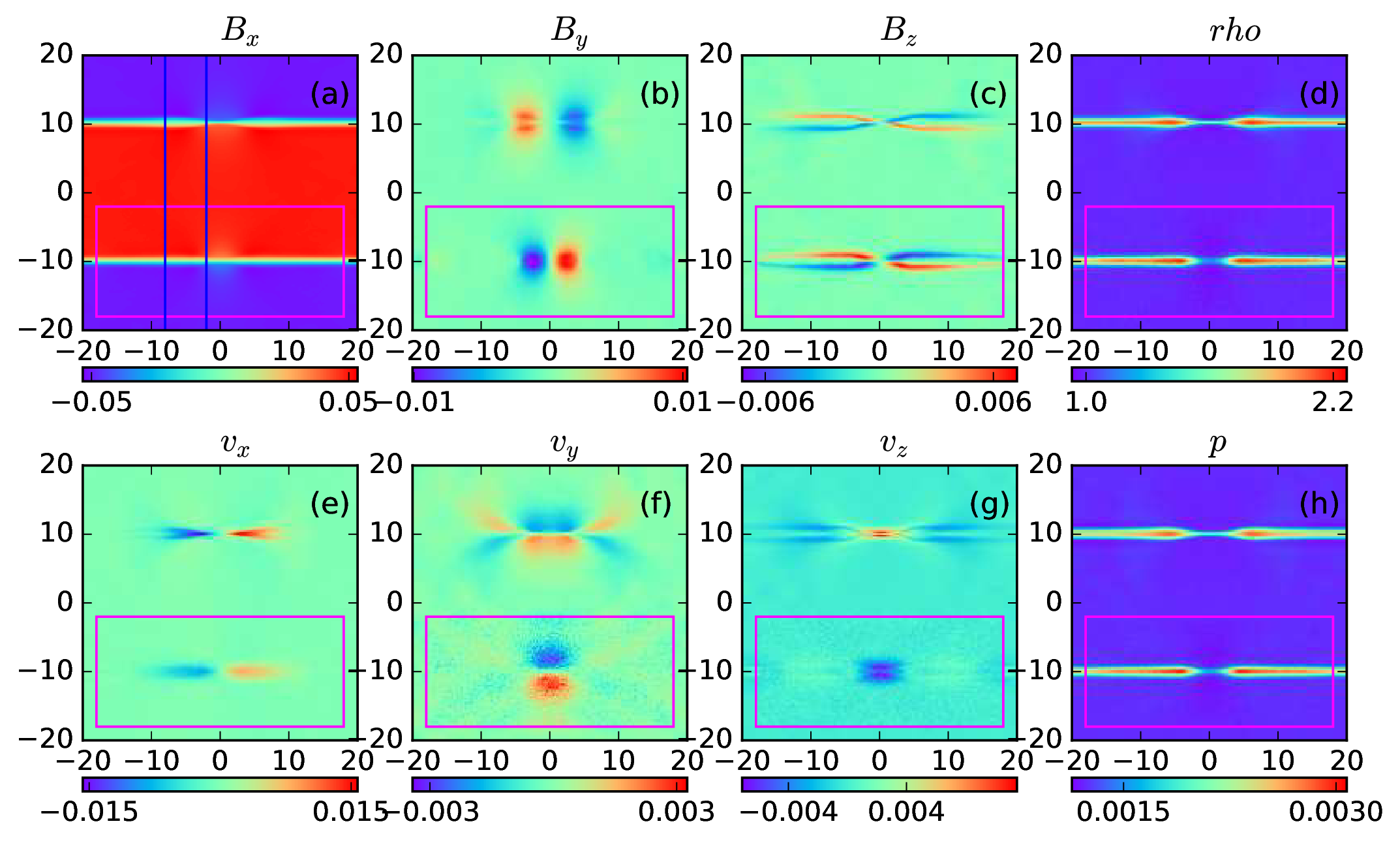}
\caption{Double GEM challenge state of two-way coupled simulation at $t=250\omega_{p,i}^{-1}$. This is coupling of iPIC3D with Hall-MHD. The magenta box shows the PIC domain, and the two vertical lines show where the slices are taken, which are shown in Fig.~\ref{DGEM_line}.}
\label{DGEM_color}
\end{figure}

\begin{figure}[h]
\centering
\includegraphics[width=1.0\textwidth]{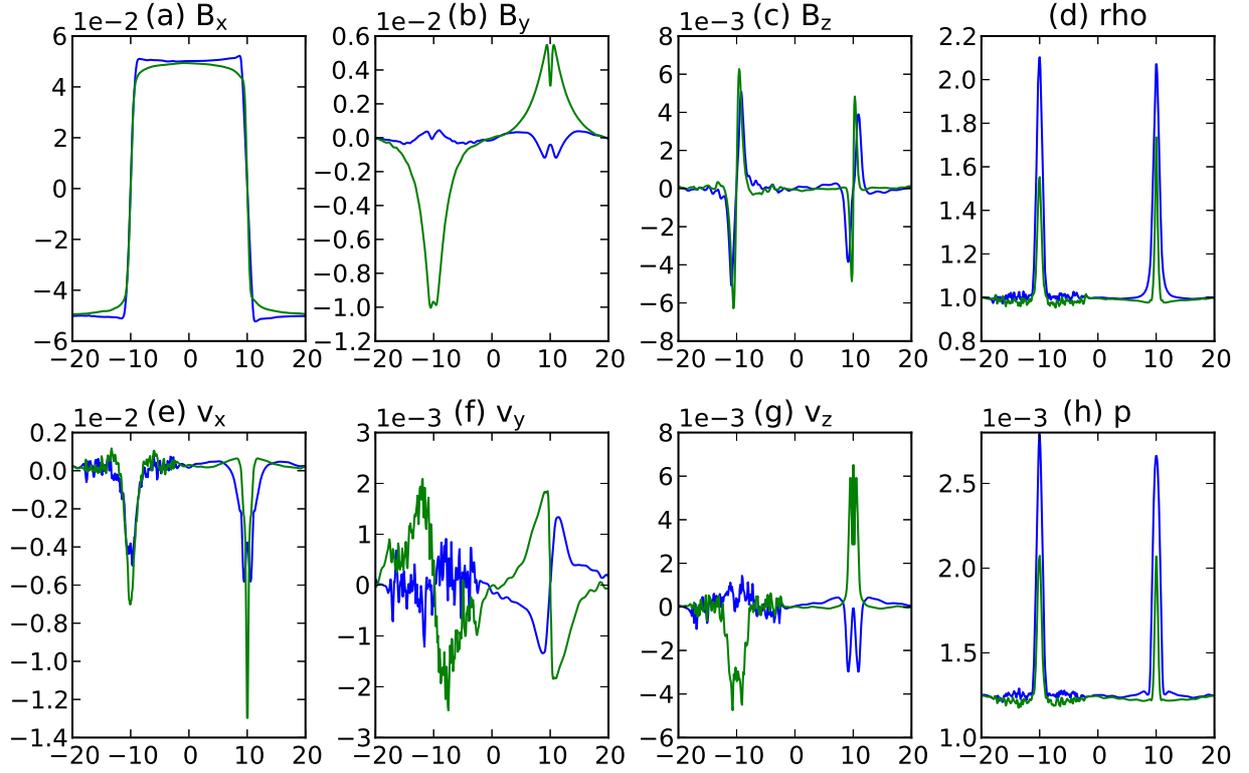}
\caption{Cuts along the two lines shown in Fig.~\ref{DGEM_color}. The blue line is the cut at $x=-8$ and the green line is at $x=-2$. The $x$-axis shows the $y$-coordinate along the cut. The left-hand-side is the two-way coupled solution, the right-hand-side is the Hall-MHD solution.}
\label{DGEM_line}
\end{figure}

\begin{figure}[h]
\centering
\includegraphics[width=1.0\textwidth]{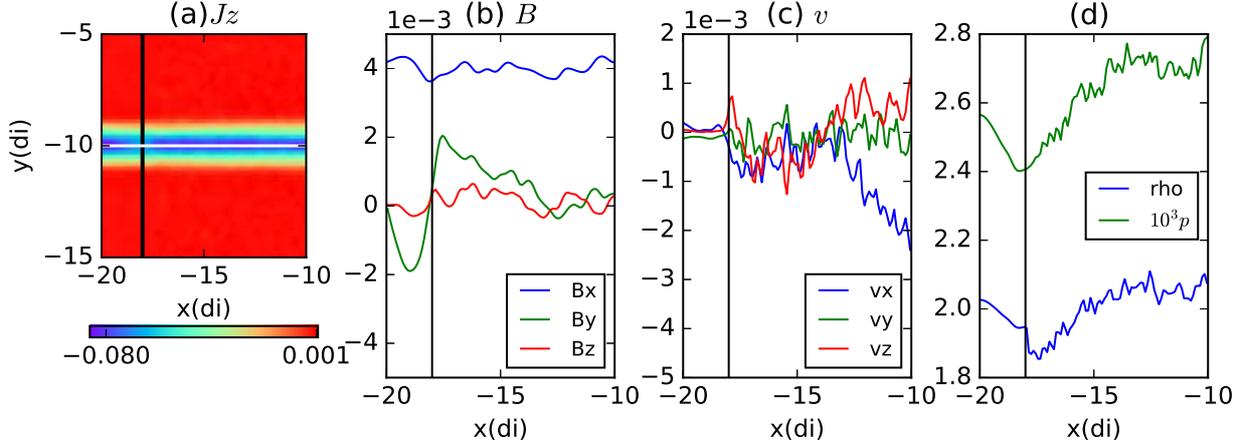}
\caption{{ (a) The out-of-plane current density $J_z$ for the simulation shown in Fig.~\ref{DGEM_color}. This is zoomed into the box $-20\le x \le -10$ and $-15 \le y \le -5$. The vertical black line indicates the MHD-PIC interface, with MHD on the left and PIC feedback on the right. Traces of MHD quantities are taken along the white horizontal line at $y=-10$ which lies along the current sheet and cuts the MHD-PIC interface at $x=-18$. (b) shows traces of the magnetic field components, (c) shows the traces of velocity components, and (d) shows traces of density and pressure.}}
\label{solution_quality_at_boundary_DGEM_refreview}
\end{figure}

\begin{figure}[h]
\centering
\includegraphics[width=1.0\textwidth]{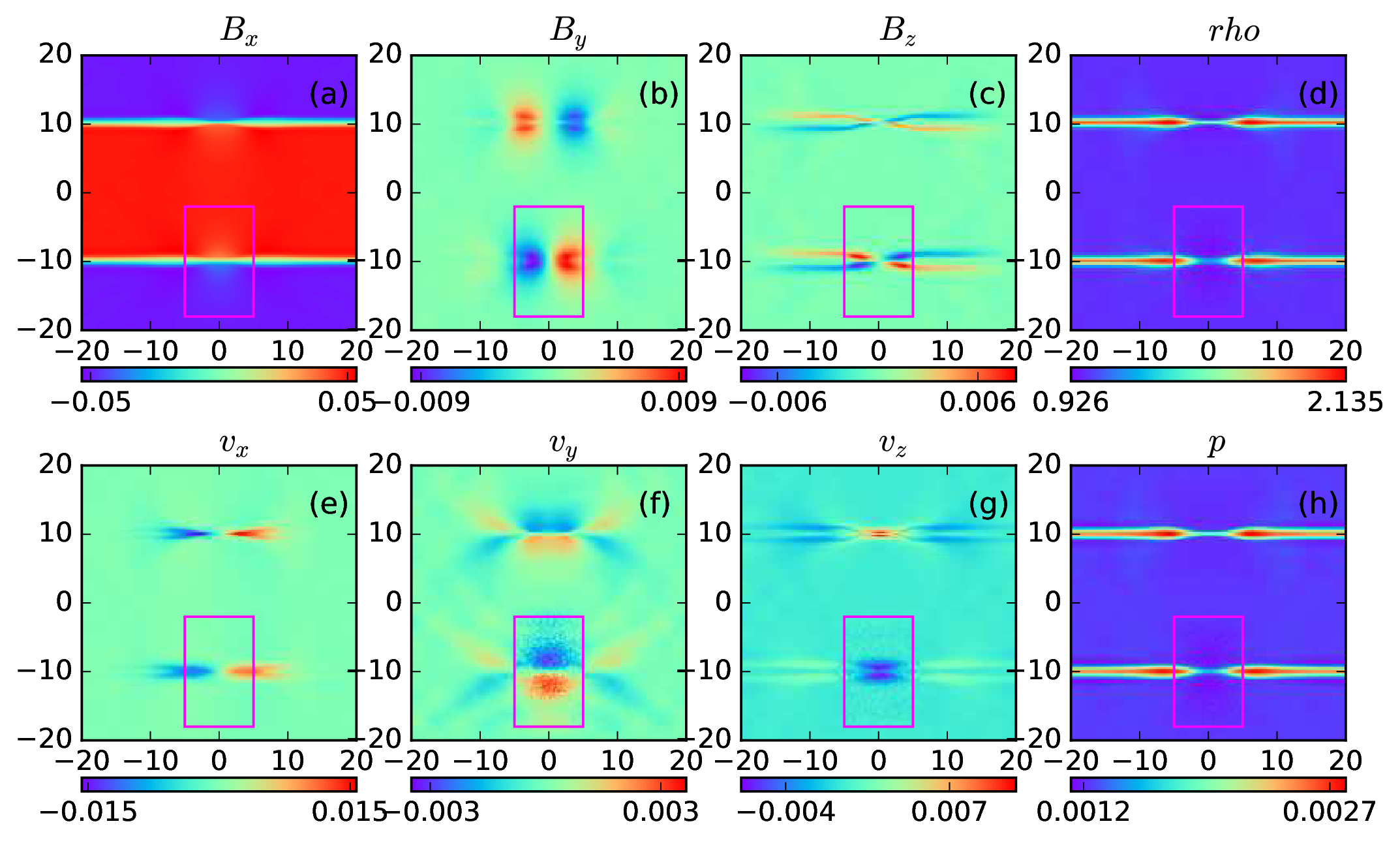}
\caption{Double GEM challenge state of two-way coupled simulation at $t=250\omega_{p,i}^{-1}$. This simulation is very similar to the one shown in Fig.~\ref{DGEM_color} except the PIC feedback is in a smaller box shown by the magenta rectangle. Outside the PIC feedback region, the solution looks more like Hall-MHD solution in the top current layer.}
\label{color_DGEM_smallbox_refreview}
\end{figure}

\begin{figure}[h]
\centering
\includegraphics[width=1.0\textwidth]{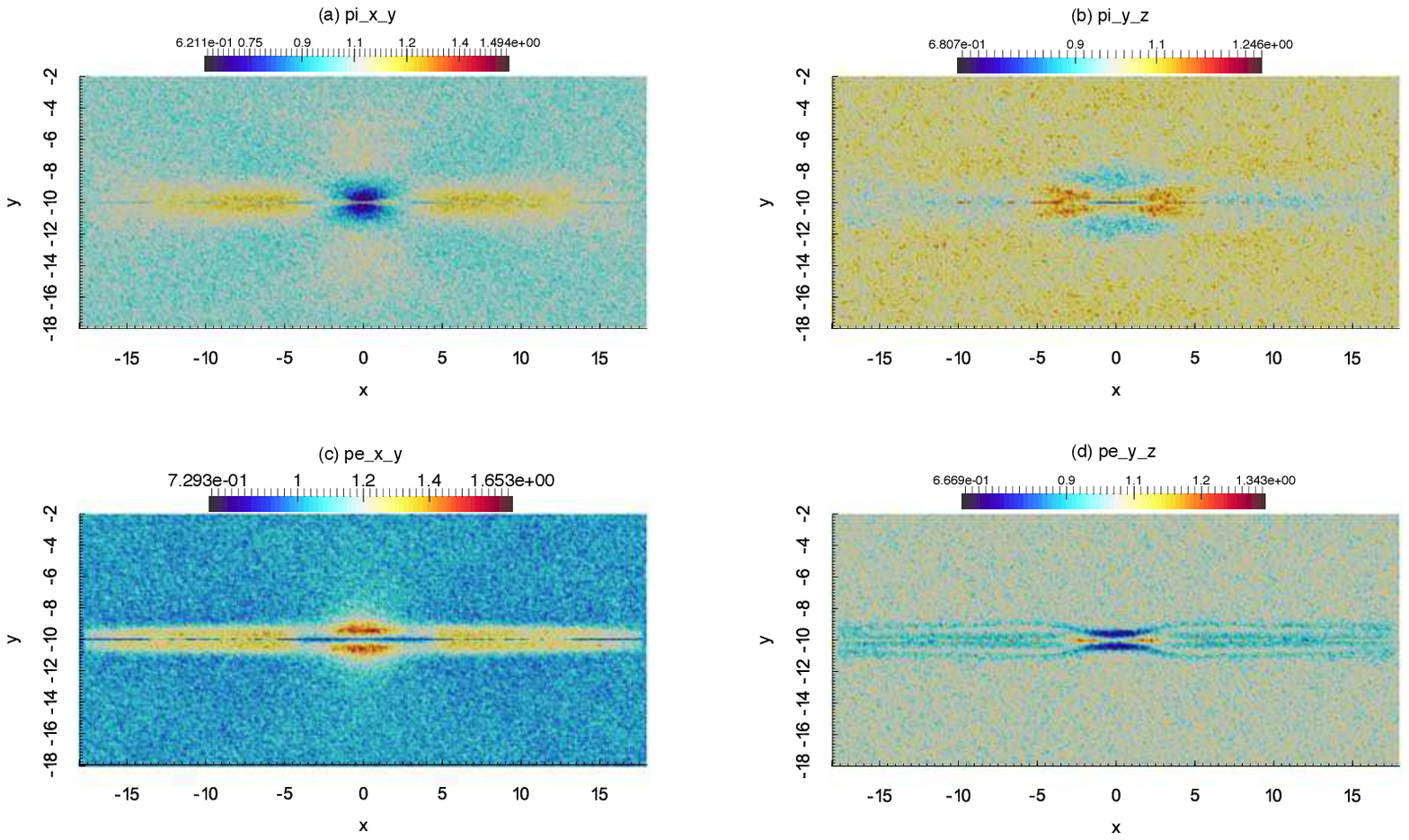}
\caption{The pressure tensor is split into parallel ($p_{\parallel}$) and perpendicular ($p_{\perp 1}$ and $p_{\perp 2}$) components as described in the text. Above figures show the PIC domain in the double GEM challenge simulation at $t=250\omega_{p,i}^{-1}$. (a) shows the parallel to perpendicular ratio for ion pressures $p_{i,\parallel}/p_{i,\perp 1}$, (b) shows the ratio between two perpendicular ion pressure components $p_{i,\perp 1}/p_{i,\perp 2}$, (c) shows the parallel to perpendicular ratio for electron pressures $p_{e,\parallel}/p_{e,\perp 1}$, and (d) shows the ratio between two perpendicular electron pressure components $p_{e,\perp 1}/p_{e,\perp 2}$.}
\label{anisotropy}
\end{figure}

The PIC simulation box is setup around the bottom current sheet within the range $-18\le x \le 18$ and $-18\le y \le -2$, with the same resolution and $360 \times 160$ cells. The width of the transition layer is $\delta = 2d_i$. The time step here is $\Delta t = 0.05\omega_{p,i}^{-1}$. The PIC simulation is started at $t=20\omega_{p,i}^{-1}$. We show the state of the two-way coupled system at time $t=250\omega_{p,i}^{-1}$ in Fig.~\ref{DGEM_color}. There is general agreement between the upper sheet which is simulated with Hall-MHD and the lower sheet which is simulated with PIC. There is no distortion produced at the boundaries. In Figs.~\ref{DGEM_color}(c) and (g), both current sheets produce the out-of-plane magnetic field $B_z$ and velocity $v_z$ which is a characteristic of Hall-physics. There are differences in the structure of the reconnection region in the pressure and density plots, where we see that the Hall MHD produces a smaller opening angle compared to the PIC solution. This is also seen in the outflow velocity $v_x$. However, GEM setups with Hall-MHD and PIC codes do show some differences in these structures.

We take two line-cuts through the simulation domain at $x=-8$ (farther from the X-point) and $x=-2$ (closer to the X-point) as shown by the two lines in Fig.~\ref{DGEM_color}(a). The traces of the various quantities along these cuts are shown in Fig.~\ref{DGEM_line}. The negative part (left side) of the $x$-axis represents the current sheet with the two-way coupled solution, whereas the positive part (right side) shows the Hall-MHD solution. Clearly we can see more noise in the PIC feedback part. The magnetic field perturbation ($B_y$) is stronger in the PIC part compared to Hall-MHD. The outflow velocity ($v_x$) is comparable between PIC and Hall-MHD farther away from the X-point, but closer to the X-point Hall-MHD shows a stronger outflow. The $v_y$ is very similar between the two cases. We do not see any distortion or discontinuity at the boundary of the PIC region at $y=-18$ and $y=-2$. 

\begin{figure}[h]
\centering
\includegraphics[width=1.0\textwidth]{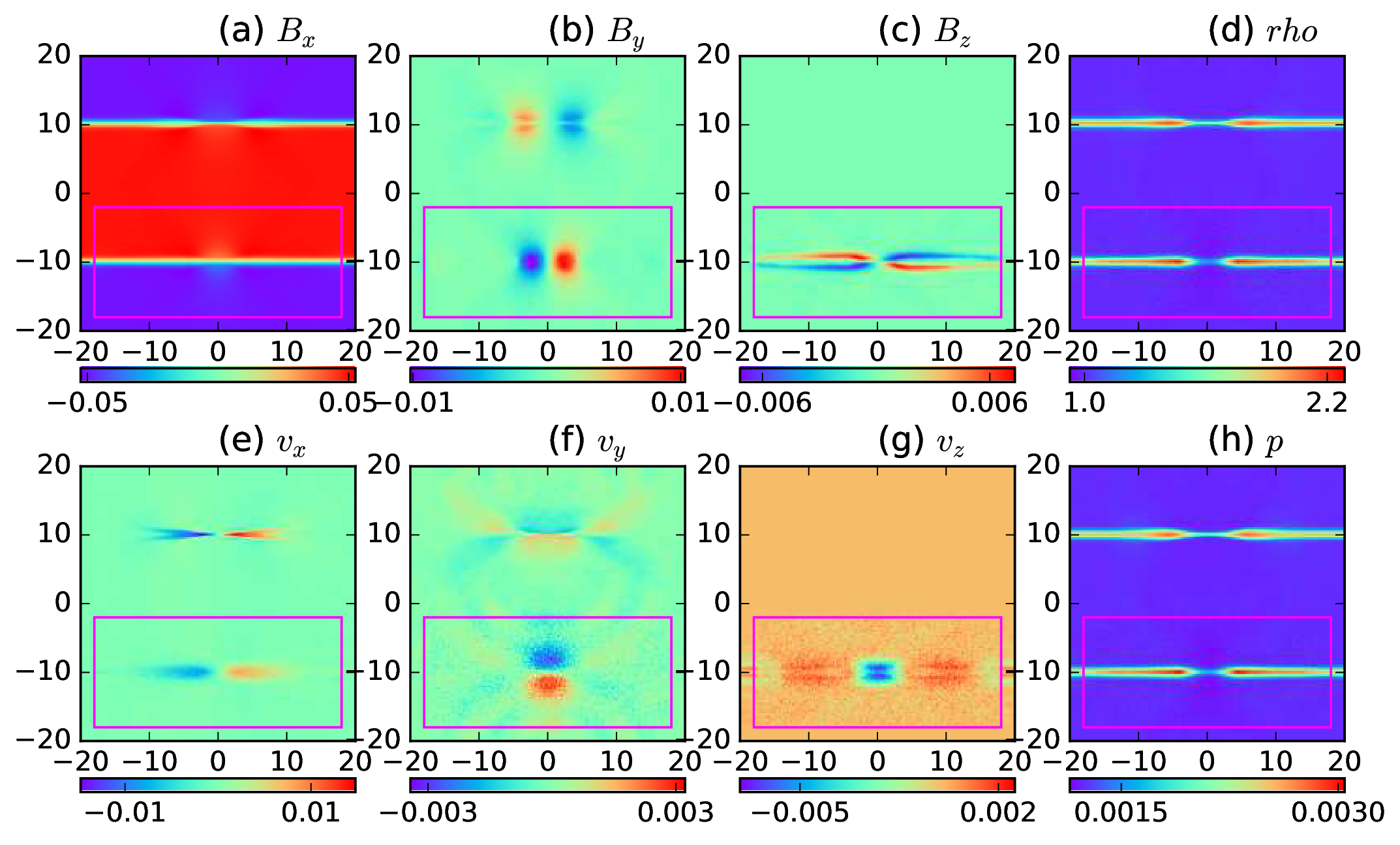}
\caption{Double GEM challenge done by coupling PIC with MHD, not Hall-MHD. State of two-way coupled simulation at $t=250\omega_{p,i}^{-1}$. The magenta box shows the region of PIC feedback.}
\label{DGEM_onlyMHD_color}
\end{figure}

{We want to check the quality of the solution near the MHD-PIC interface region in Fig.~\ref{DGEM_color} where it cuts through the current sheet. For this we zoom into the small box around it, with range $-20\le x \le -10$ and $-15 \le y \le -5$. The out-of-plane current density ($J_z$) profile in this box is shown in Fig.~\ref{solution_quality_at_boundary_DGEM_refreview}(a). The current sheet center lies along the $y=-10$ line, while the MHD-PIC interface lies along the $x=-18$ line, indicated by the black vertical line, with the MHD domain lying to the left of this line, while the PIC feedback applies to the right of it. We see that the current density shows a slight dip at this interface, but it is reasonably well-behaved. We take a trace of all the MHD quantities along the current sheet at $y=-10$ line (indicated by the horizontal white line). The quantities are plotted along $x$ in the other figures. Fig.~\ref{solution_quality_at_boundary_DGEM_refreview}(b) shows the magnetic field components along this line, and the MHD-PIC interface lies at the $x=-18$ position, indicated by the black vertical line. We find that $B_x$ and $B_z$ do not show any jump at the MHD-PIC interface. $B_y$ shows a slight fluctuation at this interface. We have found that fixing the magnetic field at the three outermost cell-centers, as discussed in Sec.~\ref{MHDtoPIC} slightly reduces this problem compared to fixing the $\mathbf{B}$ field only in one cell. Fig.~\ref{solution_quality_at_boundary_DGEM_refreview}(c) shows the velocity components along the current sheet as it crosses MHD-PIC interface at $x=-18$. All the three components are well-behaved at the interface, except for the introduction of particle noise inside the PIC domain. Fig.~\ref{solution_quality_at_boundary_DGEM_refreview}(d) shows the quantities of density and pressure, with the pressure multiplied by $10^3$ for ease of viewing. The pressure is smooth across the boundary, while the density shows a jump of less than $5\%$ across the MHD-PIC interface. The other side of the MHD-PIC interface at the current sheet shows similar behavior.}

In this case, the PIC box is taken quite large so that its boundaries are far away from the diffusion region. { We have tried a simulation with a PIC box with smaller extent in the X-direction, $-5 d_i\le x \le 5 d_i$. The state of this simulation again after $t=250\omega_{p,i}^{-1}$ is shown in Fig.~\ref{color_DGEM_smallbox_refreview}. This case also shows satisfactory results and can be compared with Fig.~\ref{DGEM_color}. However, we observe changes across the interface boundary. Comparing Fig.~\ref{color_DGEM_smallbox_refreview}(b) with Fig.~\ref{DGEM_color}(b) the $B_y$ perturbation with a smaller PIC-feedback region is different compared to the $B_y$ perturbation with a larger PIC-feedback domain. Comparing the quadrupolar $B_z$ field of Fig.~\ref{color_DGEM_smallbox_refreview}(c) with Fig.~\ref{DGEM_color}(c), we see that inside the smaller PIC domain they match well. However, outside of it, the field resembles more like the Hall-MHD field in the top current layer, which is slightly weaker and has a slightly different shape compared to the field in the larger PIC feedback region. Similarly in the outflow velocity $v_x$ (Figs.~\ref{color_DGEM_smallbox_refreview}(e) and ~\ref{DGEM_color}(e)) and vertical velocity $v_y$ (Figs.~\ref{color_DGEM_smallbox_refreview}(g) and ~\ref{DGEM_color}(g)), the solution of the smaller PIC domain simulation matches with the solution of the larger PIC domain simulation, within the small domain. However, outside of it, the solution resembles more of the Hall-MHD solution rather than the larger PIC domain solution. Thus, it does not seem that PIC feedback in a localized region smaller than the span of the quadrupolar magnetic field can modify the Hall-MHD solution outside to resemble a global PIC solution.  This indicates that we should take a PIC domain which at least covers the quadrupolar magnetic field structure.}

We have seen that the embedded PIC solution qualitatively resembles the Hall-MHD solution. We are interested in finding whether there is something purely kinetic in the embedded PIC domain which Hall-MHD cannot reproduce. We can expect such a signature in pressure anisotropy and agyrotropy~\cite{CazzolaInnocenti2016}. Fig.~\ref{anisotropy} shows a measure of these quantities. The pressure tensor is split into parallel ($p_{\parallel}$ along the unit local magnetic field, $\hat{b}$) and perpendicular components ($p_{\perp 1}$ along $\hat{b}\times\hat{z}$ direction and $p_{\perp 2}$ along $(\hat{b}\times\hat{z})\times\hat{b}$) direction. Pressure anisotropy is the ratio $p_{\parallel}/p_{\perp 1}$, shown in Figs.~\ref{anisotropy}(a) and (c) for the ions and electrons respectively. We see that this ratio is different from unity, showing strong anisotropy for both species close to the current sheet and the X-point. To handle this we would need anisotropic pressure tensor in MHD simulations. Furthermore, pressure agyrotropy is the difference between two perpendicular pressure components. We show the ratio of $p_{\perp 1}/p_{\perp 2}$ for the two species in Figs.~\ref{anisotropy}(b) and (d). This ratio is also different from unity close to the current sheet and the X-point, showing agyrotropy. This cannot be handled by MHD since this comes from anisotropy along the gyromotion of particles. This is a purely kinetic effect which will not be captured in MHD. Furthermore, the anisotropy and agyrotropy for the electrons extends throughout the PIC domain, touching the MHD-PIC boundary also. Thus, the embedded PIC domain is producing purely kinetic effects which Hall-MHD with anisotropic pressure cannot capture.

We have done the same simulation as above but coupled with MHD instead of Hall-MHD. This can be simply done by setting the parameter $\eta_h=0$. The results from this simulation are shown in Fig.~\ref{DGEM_onlyMHD_color}. The advantage of this is that the time step can be relaxed, which we take to be $\Delta t = 0.1\omega_{p,i}^{-1}$, which is double of that taken in the coupling with Hall-MHD. In this case also, the PIC domain produces a very similar solution comparing to Figs.~\ref{DGEM_color}. On the other hand, the MHD solution is unable to produce either the out-of-plane magnetic field $B_z$ or velocity $v_z$. Moreover in Fig.~\ref{DGEM_onlyMHD_color}(f) we can see waves emanating from the X-point that propagate smoothly from the PIC domain to the MHD domain. This is encouraging since we can now also give PIC feedback to an MHD simulation of reconnection and correctly reproduce the kinetic physics.

\section{Summary and discussion}
This work shows the first results of coupling the MHD code MPI-AMRVAC with the semi-implicit PIC code iPIC3D. { It follows a similar philosophy as an earlier work of coupling iPIC3D with BATS-R-US~\cite{DaldorffToth2014}. This work utilizes a larger boundary interface zone for the magnetic fields as well as particles. This coupling utilizes a weight function in the PIC to MHD coupling in the transition layer to smoothly transition from the PIC domain to MHD domain. We do not need to make an adjustment to the charge density in order to keep the solution stable at the boundaries. The time stepping scheme utilized is second order accurate in time, with time averaged boundary conditions passed to PIC. When the MHD time step is an integer multiple of the PIC time step, then the boundary condition for each individual PIC step is obtained by linear interpolation between the start and end of MHD time step. We have coupled PIC with a scalar MHD pressure instead of anisotropic pressure tensor, which however still seems to work very well. The parallelization strategy followed is to utilize the maximum number of processors available for both MHD and PIC, with each processors alternately processing the MHD and PIC simulations. This leads to increased communication costs which has to be improved in the future.}

This work shows the physics viability of such a coupling scheme. The physics at small scales is correctly reproduced by the PIC code. The simulation of an Alfv{\'e}n wave is difficult as it is a long-wavelength solution of the plasma dispersion relation, which is not correctly reproduced in PIC, and we suspect larger kinetic simulation domains will be required to reproduce the MHD behavior. Whistler waves on the other hand are well-reproduced when we couple PIC to Hall-MHD, as the whistler wave is a solution of the Hall-MHD equations and also has some kinetic physics in it. The fast magnetosonic wave is handled well by the coupling even though we use only a scalar pressure in the MHD equations. The fast wave steepening is also handled well by this coupling, indicating that it might be suitable for shock problems. The GEM-like setup also behaves well and shows expected results. The PIC domain has to be large enough to capture all the kinetic signatures of magnetic reconnection. However, the PIC domain is able to reproduce this magnetic field when coupling with both Hall-MHD and MHD. This is encouraging for problems where reconnection is closely coupled with large scales, like stellar flares and magnetospheres.

The major goal of this work is to be able to simulate large systems where kinetic effects are localized but affect global dynamics. In order to truly develop this potential, the coupling and data exchange methodology will have to be optimized. Moreover, this coupling can also be viewed as a way for providing open boundary conditions to a PIC simulation, as we know the MHD solution outside.  Also now it is very easy to setup a PIC simulation by just specifying the initial MHD state. These features are very general, meaning a vast variety of physical setups can be simulated with this method.

\section{Acknowledgement}
This research was supported by the Interuniversity Attraction Poles
Programme (initiated by the Belgian Science Policy Office, IAP P7/08 CHARM) and by the KU Leuven GOA/2015-014. Simulations used the VSC (Flemish Supercomputer Center) funded by the Hercules foundation and the Flemish government. KM would like to thank Maria Elena Innocenti for useful discussions.

%

\section*{References}

\bibliography{mybibfile}

\end{document}